\pdfoutput=1
\newcommand*{\ATLASLATEXPATH}{latex/}

\documentclass[cernpreprint,texlive=2011,txfonts,UKenglish]{latex/atlasdoc}

\usepackage{\ATLASLATEXPATH atlaspackage}
\usepackage{\ATLASLATEXPATH atlasbiblatex}

\usepackage{\ATLASLATEXPATH atlasphysics}
\usepackage{multirow}
\usepackage{units}
\usepackage{dingbat}
\usepackage{bm}
\usepackage{booktabs,siunitx}
\graphicspath{{logos/}{figures/}}

\usepackage{haa4b_paper-defs}

\AtlasTitle{Search for the Higgs boson produced in association with a
$W$ boson and decaying to four $b$-quarks via two spin-zero particles 
in $pp$ collisions at \unit[13]{\TeV} with the ATLAS detector} 

\author{The ATLAS Collaboration}

\AtlasRefCode{GROUP-2016-01}

\PreprintIdNumber{CERN-EP-2016-135}

\AtlasDate{\today}

\AtlasJournal{EPJC}
\AtlasJournalRef{\EPJC 76 (2016) 605}
\AtlasDOI{10.1140/epjc/s10052-016-4418-9}

\AtlasAbstract{%
This paper presents a dedicated search for exotic decays of the Higgs boson to a
pair of new spin-zero particles, $H \rightarrow aa$, where the
particle $a$ decays to $b$-quarks and has a mass in the range of
\unit[20--60]{\GeV}. The search is performed
in events where the Higgs boson is produced in association with
a $\Wboson$ boson, giving rise to a signature of a lepton (electron or
muon), missing transverse momentum, and multiple jets from $b$-quark
decays. The analysis is based on the full dataset of $pp$ collisions
at $\sqrt{s} = \unit[13]{\TeV}$ recorded in 2015 by the ATLAS detector at the CERN Large Hadron Collider,
corresponding to an integrated luminosity of \unit[3.2]{$\ifb$}. No
significant excess of events above the Standard Model prediction is 
observed, and a $95\%$ confidence-level upper limit is derived for the 
product of the production cross section for $pp \to WH$ times the
branching ratio for the decay $H \to aa \to 4b$. The upper limit
ranges from 6.2~pb for an $a$-boson mass $m_a = \unit[20]{\GeV}$ to
1.5~pb for $m_a = \unit[60]{\GeV}$.
}

\hypersetup{pdftitle={ATLAS document},pdfauthor={The ATLAS Collaboration}}

\begin{document}

\maketitle



\section{Introduction}
\label{sec:intro}

Following the discovery of the Higgs boson by the ATLAS and
CMS Collaborations~\cite{ATLAS_Higgs,CMS_Higgs} at the Large Hadron
Collider (LHC), a comprehensive programme of measurements of the
properties of this particle is underway. These measurements could uncover
deviations from expected Standard Model (SM) branching ratios or allow 
for the possibility of decays into non-SM particles. Existing
measurements constrain the non-SM or ``exotic'' branching ratio of the 
Higgs boson decays to less than approximately $30\%$ at $95\%$ confidence level 
(CL)~\cite{Aad:2015gba,Khachatryan:2014jba,Khachatryan:2016vau}. 
Exotic decays are predicted by many theories of physics beyond the
SM~\cite{Curtin:2013fra}, including those with an extended Higgs sector
such as the Next-to-Minimal Supersymmetric Standard Model 
(NMSSM)~\cite{Dobrescu:2000yn,Ellwanger:2003jt,Dermisek:2005ar,Chang:2008cw,Morrissey:2008gm},
several models of dark
matter~\cite{Silveira:1985rk,Pospelov:2007mp,Draper:2010ew,Ipek:2014gua,Martin:2014sxa}, 
models with a first-order electroweak phase
transition~\cite{Profumo:2007wc,Blinov:2015sna}, and theories 
with neutral naturalness~\cite{Burdman:2006tz,Craig:2015pha,Curtin:2015fna}.

One of the simplest possibilities is that the Higgs boson
decays to a pair of new spin-zero particles, $a$, which in turn
decay to a pair of SM particles, mainly fermions~\cite{Curtin:2013fra}.\footnote{Throughout this paper, the symbol
  for a particle may be used to represent both the particle and its
  antiparticle.} These kinds of models have been used
to explain the recent observations of a gamma-ray excess from the galactic 
centre by the Fermi Large Area Telescope
(FermiLAT)~\cite{TheFermi-LAT:2015kwa,Boehm:2014hva}. 
Several searches have been performed for $H\to aa$. The D0 and ATLAS 
Collaborations have searched for a signal of $H\to aa \to 2\mu
2\tau$ in the $a$-boson mass ranges $\unit[3.7]{\GeV} \leq m_a \leq
\unit[19]{\GeV}$~and~$\unit[3.7]{\GeV} \leq m_a \leq \unit[50]{\GeV}$, 
respectively~\cite{Abazov:2009yi,Aad:2015oqa}. The D0 and CMS
Collaborations have searched for the signature $H \to aa \to
4\mu$ in the range $2 m_{\mu} \leq m_a \leq 2
m_{\tau}$~\cite{Abazov:2009yi,Khachatryan:2015wka}. 
In this analysis, the $a$-boson is assumed to have a negligibly small lifetime. Several
other searches have been performed by the ATLAS, CMS and LHCb Collaborations 
for signatures that may correspond to a long-lived $a$-boson:
displaced decays of jets or displaced decays of collimated 
leptons~\cite{Aad:2014yea, CMS:2014hka, Aad:2015uaa, Aad:2015asa, CMS:2014wda, Aaij:2014nma}.

The result presented in this paper covers an unexplored decay mode in
searches for $H \to aa$ by considering $a\to bb$. The $a$-boson can be 
either a scalar or a pseudoscalar under parity transformations, since
the decay mode considered in this search is not sensitive to the
difference in coupling. An example
of a model with predominant $a\to bb$ decays is one where the new scalar
mixes with the SM Higgs boson and inherits its Yukawa couplings~\cite{Curtin:2013fra}.
This search focuses on the $pp \to WH$ process, with
$W\to\ell\nu$  ($\ell=e,\mu$) and $H \to  2a \to 4b$ in the
range $\unit[20]{\GeV}<m_a<\unit[60]{\GeV}$. 
The resulting signature has a single lepton accompanied by a high 
multiplicity of jets originating from a bottom quark ($b$-jets). Since
the $b$-jets are produced from the decay of the Higgs boson, they tend to 
have low transverse momentum ($\pt$) compared to $m_H$ and can 
be overlapping, especially for light $a$-bosons. Events with an
electron or muon, including those produced via leptonically 
decaying $\tau$-leptons, are considered. The $WH$ process is chosen
for this search because the charged lepton in the final state 
provides a powerful handle to efficiently trigger and 
identify these events against the more ubiquitous background process 
of strong production of four $b$-jets.
Several kinematic variables, including the reconstructed masses in 
the decay $H\to 2a \to 4b$, are used to identify signal events. The 
background estimation techniques, systematic uncertainties and 
statistical treatment closely follow those used in other ATLAS
searches with similar 
signatures~\cite{Aad:2015gra,Aad:2015kqa,Aad:2015pja,Aad:2015typ}.


\section{ATLAS detector}
\label{sec:detector}

The ATLAS detector~\cite{ATLASJINST} covers nearly the entire solid angle\footnote{ATLAS 
uses a right-handed coordinate system with its origin at the nominal interaction point (IP) in the 
centre of the detector and the $z$-axis coinciding with the axis of the beam pipe.  The $x$-axis points from
the IP to the centre of the LHC ring, and the $y$-axis points upward. Cylindrical coordinates ($r$,$\phi$) are used 
in the transverse plane, $\phi$ being the azimuthal angle around the beam pipe. The pseudorapidity is defined in 
terms of the polar angle $\theta$ as $\eta = - \ln \tan(\theta/2)$.} around the collision point.
It consists of an inner tracking detector surrounded by a thin superconducting solenoid magnet producing a
2~T axial magnetic field, electromagnetic and hadronic calorimeters, and an external muon spectrometer
incorporating three large toroid magnet assemblies. The inner detector consists of a high-granularity silicon pixel detector, 
including the newly installed insertable B-layer~\cite{IBL}, and a silicon microstrip tracker,
together providing precision tracking in the pseudorapidity range $|\eta|<2.5$, complemented by a
transition radiation tracker providing tracking and electron identification information for $|\eta|<2.0$.
The electromagnetic (EM) sampling calorimeter uses lead as the absorber material 
and liquid argon (LAr) as the active medium, and is divided into barrel ($|\eta|<1.475$) and end-cap ($1.375<|\eta|<3.2$) regions.  
Hadron calorimetry is also based on the sampling technique, with
either scintillator tiles or LAr as the active medium, and with 
steel, copper, or tungsten as the absorber material. The scintillator
tile calorimeter is divided into barrel ($|\eta|<1.0$) and end-cap
($0.8<|\eta|<1.7$) regions, and the LAr hadronic calorimeter includes
an end-cap ($1.5<|\eta|<3.2$) and a forward ($3.1<|\eta|<4.9$) region.
The muon spectrometer measures the deflection 
of muons with $|\eta|<2.7$ using multiple layers of high-precision tracking chambers in a toroidal field of 
approximately 0.5~T and 1~T in the central and end-cap regions of ATLAS, respectively. The muon spectrometer is also 
instrumented with separate trigger chambers covering $|\eta|<2.4$.
A two-level trigger system, consisting of a custom-hardware level
followed by a software-based level, is used to reduce the event rate to a maximum 
of around 1 kHz for offline storage~\cite{ATL-DAQ-PUB-2016-001}.


\section{Event samples and object selection}
\label{sec:data_sim_obj}

The search presented in this paper is based on the proton--proton
($pp$) collision dataset collected by the ATLAS 
detector at the LHC
at $\sqrt{s}=\unit[13]{\TeV}$ with 25~ns bunch spacing during 2015. 
The full dataset corresponds to an integrated luminosity of
$\unit[3.2]{\ifb}$. 
The data for this search were collected using the single-electron
or single-muon triggers with the lowest transverse momentum thresholds 
available~\cite{ATL-DAQ-PUB-2016-001}. 

Electron candidates are reconstructed by associating an inner-detector
track with an isolated energy deposit in the EM
calorimeter~\cite{Aad:2014fxa,ATLAS-CONF-2014-032}. 
Candidates are identified using the tight quality criteria and are
required to have $\pt>\unit[25]{\GeV}$ and 
$|\eta|<2.47$, excluding the transition region between the barrel and end-cap
EM calorimeters, $1.37<|\eta|<1.52$. Muon candidates are
reconstructed by combining matching tracks in the inner detector and
the muon spectrometer~\cite{muons13TeV}, and are required to satisfy 
the medium quality criteria and to have $\pt>\unit[25]{\GeV}$ and $|\eta|<2.4$. 
Events are required to have exactly one reconstructed electron or muon
that is matched within a cone of size 
$\Delta R \equiv \sqrt{(\Delta \eta)^2 + (\Delta \phi)^2} = 0.15$ to the
lepton candidate reconstructed by the trigger algorithms. 

In order to distinguish leptons produced in the decays of $\Wboson$
bosons from those produced in the decays of heavy-flavour hadrons, all 
lepton candidates are required to be consistent with originating from
the primary interaction vertex, chosen as the vertex with the highest 
sum of the $\pt^2$ of its associated tracks. 
Furthermore, since lepton candidates arising from background sources, 
such as decays of hadrons, are typically embedded in jets, all lepton
candidates are required to be isolated from other particles in the
event. This is achieved by imposing a maximal allowed value on the
energy deposited in the calorimeter and/or the momentum of inner-detector tracks within a cone of
$\Delta R = 0.2$ around the direction of the lepton candidate's momentum. 
The isolation criteria, dependent on \pt\ and $\eta$, are
applied to produce a nominal efficiency of at least $90\%$ for 
electrons and muons from $Z\to \ell^+\ell^-$ decays after all other
identification criteria are applied~\cite{muons13TeV}.  

Jets are reconstructed from clusters~\cite{Aad:2016upy} of energy in the
calorimeters using the anti-$k_{t}$ clustering algorithm~\cite{antikt,antikt2} 
with radius parameter $R=0.4$. Jets are required to have $\pt >
\unit[20]{\GeV}$ and $|\eta|<2.5$, and they are calibrated using energy- and
$\eta$-dependent corrections.
A track-based veto is used to suppress contributions from jets arising from
additional $pp$ interactions (pile-up)~\cite{Aad:2015ina}.
Jets consistent with the hadronisation of a $b$-quark are identified using 
information from track impact parameters and secondary vertices,
which are combined in a multivariate discriminant~\cite{ATL-PHYS-PUB-2015-022}. The
efficiency to identify $b$-quark jets ($b$-tagging) is approximately
$77\%$ for a factor of 126 in rejection against light-quark and gluon
jets, about 5 against jets originating from $c$-quarks, and about
10 against hadronically decaying $\tau$-leptons, as determined in a simulated sample of
top-quark pair ($\ttbar$) events~\cite{Aad:2015ydr,btag-Run2,ATL-PHYS-PUB-2015-022}. 
Jets tagged by this multivariate discriminant, independently of the
flavour of the quark that initiated it, are called $b$-tagged jets
throughout the text, while the term $b$-jet is reserved for those jets
originating from $b$-quark decays, as determined from simulation.

Jets are required to be separated from the lepton candidates by
$\Delta R$ larger than 0.2 or 0.4 for electrons or muons,
respectively. Electrons separated from the nearest jet by $0.2<\Delta
R < 0.4$ are considered part of the jet and not a lepton candidate. 
The transverse momentum imbalance
$\ensuremath{\vec{E}_{\text{T}}^{\text{miss}}}$, the magnitude of which 
($\met$) is commonly referred to as missing transverse momentum, is 
defined as the negative vector sum of the transverse momenta of
calibrated selected objects, such as electrons, muons and
jets. The transverse momenta of charged-particle tracks compatible 
with the primary vertex and not matched to any of those objects are 
also included in the negative vector sum~\cite{Aad:2012re,ATL-PHYS-PUB-2015-027}.

\section{Signal and background modelling}
\label{sec:bkg_modeling}

Simulated event samples are used to study the
characteristics of the signal and to calculate its acceptance, and
are also used for most of the SM background estimation. 
Signal samples of associated Higgs boson production with a $\Wboson$
boson, $pp \to W H$, are generated with \powheg\ v2--r3033~\cite{Frixione:2007nw,Nason:2004rx,Frixione:2007vw,Alioli:2010xd} 
using the CT10 parton distribution functions (PDFs)~\cite{Lai:2010vv}
at next-to-leading order
(NLO). A Higgs boson mass of $m_H = \unit[125]{\GeV}$ is assumed and the sample is
normalised to the next-to-next-to-leading-order (NNLO) cross section
recommended by the Higgs cross-section working
group $\sigma_{\rm SM}(WH) = 1.37$~pb~\cite{Heinemeyer:2013tqa}. 
The Higgs boson decay to two spin-zero $a$-bosons and the subsequent decay of each $a$-boson to a pair of
$b$-quarks are simulated with \pythia~v8.186~\cite{pythia8}.
The $a$-boson decay is done in the narrow-width
approximation and the coupling to the $b$-quarks is assumed to be that of
a pseudoscalar. However, since the polarisation of the quarks is not
observable, this search is insensitive to the specific parity
hypothesis. \pythia~v8.186 is used for the showering,
hadronisation, and underlying-event (UE) simulation with the A14
set of tuned parameters (tune)~\cite{ATL-PHYS-PUB-2014-021}. 
The mass of the $a$-boson is varied for different signal hypotheses in the range $\unit[20]{\GeV} \leq
m_a \leq \unit[60]{\GeV}$, in $\unit[10]{\GeV}$ mass steps. Different branching-ratio
hypotheses are obtained by scaling the signal sample normalisation.

Samples of $\ttbar$ are
also produced using the NLO \powheg\ v2--r3026 generator with the CT10
PDFs. A top-quark mass ($m_t$) of $172.5\gev$ is assumed. The
{\powheg} model parameter $h_{\rm damp}$, which controls matrix
element to parton shower (PS) matching and effectively
regulates the high-$\pt$ radiation, is set to $h_{\rm
  damp} = m_t $. This setting was found to best describe the
$\ttbar$-system $\pt$ at $\sqrt{s} = 7\tev$~\cite{ATL-PHYS-PUB-2015-002}. 
The baseline  $\ttbar$ sample is interfaced to
\pythia~v6.428~\cite{pythia6} with the Perugia 2012
tune~\cite{Skands:2010ak}. Alternative $\ttbar$ samples are generated 
using {\powheg} v2--r3026 interfaced to {\sc Herwig++} v2.7~\cite{Bahr:2008pv}
or {\amcatnlo}~\cite{Alwall:2014hca} interfaced to {\sc Herwig++}.
The effects of initial- and final-state
radiation (ISR/FSR) are explored using two alternative
{\powheg~v2--r3026}+{\pythia~v6.428} samples. The first has $h_{\rm damp}$ set to $2
m_t$, the renormalisation and factorisation scales set to half the
nominal value and uses the Perugia 2012 radHi UE tune, giving more
radiation. The second sample uses the Perugia 2012 radLo UE tune, 
has $h_{\rm damp}=m_t$ and has the renormalisation and factorisation
scales set to twice the nominal value, giving less 
radiation~\cite{ATL-PHYS-PUB-2015-011}. The $\ttbar$ samples are 
normalised to the NNLO theoretical cross section of $832^{+46}_{-51}$~pb,
obtained with {\sc
  Top++}~v2.0~\cite{Cacciari:2011hy,Beneke:2011mq,Baernreuther:2012ws,Czakon:2012zr,Czakon:2012pz,Czakon:2013goa,Czakon:2011xx}. 

The simulated $\ttbar$ events are categorised depending on the
parton-level flavour content of additional particle jets\footnote{Particle jets are reconstructed by
  clustering stable particles, excluding muons and neutrinos, using the
  anti-$k_t$ algorithm with a radius parameter $R=0.4$. Muons and
  neutrinos are excluded to better reproduce the jet reconstruction
  based on calorimeter clusters.} not originating from the decay of
the $\ttbar$ system. Events containing
at least one additional particle jet matched to a $b$-hadron are labelled
as $\ttbb$. Events containing at least one additional particle jet
matched to a $c$-hadron and no $b$-hadron are labelled as $\ttcc$. The 
$\ttbb$ and $\ttcc$ categories are generically referred to as 
$\ttbar$+HF events (with HF standing for ``heavy flavour''). Remaining 
events are labelled $\ttbar$+light-jets (referred to as $\ttbar$+light)
and also include events with no additional particle jets. 

The associated heavy-flavour jets in $\ttbar$+HF are modelled in
{\powheg}+{\pythia} via the PS evolution and are simulated with a 
five-flavour scheme. 
The $\ttbb$ modelling is improved by reweighting the top-quark
\pt, \ttbar-system \pt, and kinematic properties of the associated
particle jets not originating from the top-quark decay~\cite{Aad:2015gra}
to agree with a $\ttbb$ sample generated at NLO with {\sc Sherpa+OpenLoops}~\cite{Gleisberg:2008ta,Cascioli:2011va}.
This {\sc Sherpa+OpenLoops} sample is simulated with the four-flavour
scheme (4FS) using {\sc Sherpa}~v2.1.1~\cite{Gleisberg:2008ta} and the CT10 PDF set.

Samples of single-top-quark backgrounds corresponding to the $Wt$ and
$s$-channel production mechanisms are generated with {\powheg}
v2--r2819~\cite{Alioli:2009je,Re:2010bp} using the CT10 PDF set. Overlaps
between the \ttbar\ and $Wt$ final states are handled using the
``diagram removal'' scheme~\cite{Frixione:2005vw}.  Samples of
$t$-channel single-top-quark events are generated using the
{\powheg}~\cite{Frederix:2012dh} NLO generator that uses the 4FS. 
The single-top-quark samples are normalised to the approximate NNLO 
theoretical cross sections~\cite{Kidonakis:2011wy,Kidonakis:2010ux,Kidonakis:2010tc}.  
The parton shower, hadronisation and underlying event are modelled 
using either {\pythia}~v6.428 with the Perugia 2012 tune or {\herwigpp}~v2.7 with
the UE-EE-5~\cite{Seymour:2013qka} tune.

Samples of $W/Z$+jets events are generated with the {\sc Sherpa} v2.1.1 generator. 
The matrix-element calculation is performed up to two partons at NLO and
up to four partons at leading order (LO) using {\sc Comix}~\cite{Gleisberg:2008fv} and {\sc
   OpenLoops}~\cite{Cascioli:2011va} and uses the CT10 PDFs.
Both the $W$+jets and $Z$+jets samples are normalised to their respective 
inclusive NNLO theoretical cross section calculated with
\textsc{FEWZ}~\cite{Anastasiou:2003ds}.

Samples of diboson production $WW/WZ/ZZ$+jets events are generated with the
NLO generator {\sc Sherpa} v2.1.1. Samples of $\ttbar + \gamma/\Wboson/\Zboson$ events,
including $\ttbar + WW$, are generated with up to two additional partons
using {\amcatnlo} and interfaced to \pythia~v8.186. Samples of $\ttbar+
H$ events are generated using {\amcatnlo} and interfaced to {\herwigpp}~v2.7.

The main signal and background samples use the {\sc
  EvtGen}~v1.2.0~\cite{evtgen} program to simulate the decay of
heavy-flavour hadrons, except for those generated with {\sherpa}. All
are then processed with the full simulation of the ATLAS
detector~\cite{Aad:2010ah} based on GEANT4~\cite{geant4}. The
alternative $\ttbar$ samples used to estimate systematic uncertainties
are based on a fast
simulation of the calorimeter response~\cite{fastsim}. 
Events are generated with pile-up that is 
simulated with \pythia~v8.186~\cite{pythia8} and are reweighted so
that the distribution of the multiplicity of pile-up interactions matches the distribution observed in the data. 
Simulated event samples are processed using the same reconstruction
algorithms and analysis chain as the data. 

As described in Section~\ref{sec:analysis_strategy_overall},
backgrounds are estimated by fitting predictions derived from 
simulation to data in several background-enriched samples. The 
only background prediction not derived from simulation is the 
multijet background, which contributes to
the selected data sample when a jet is mis-reconstructed as a lepton
and satisfies the identification criteria. In the electron
channel, it consists of non-prompt electrons from heavy-flavour decays,
from unidentified photon conversions or from jets with a high fraction
of energy deposited in the EM calorimeter. In the muon
channel, it consists of heavy-flavour decays and in-flight decays of light mesons.

The multijet background contribution is evaluated from data using the ``matrix method''~\cite{Aad:2010ey,Aad:2015kqa,ATLAS:2014ffa}, 
which uses differences between the isolation properties of background
(fake/non-prompt) leptons and signal (prompt) leptons from $W$ boson decays. 
The estimate uses a sample enriched in multijet background events 
obtained by applying the full event selection except for loosening the lepton isolation requirement. Each event 
with a lepton candidate that satisfies at least the loosened isolation 
requirement is scaled by a weight that depends on whether this 
lepton candidate also satisfies the tighter isolation requirement. The 
weights are determined from the efficiencies for fake/non-prompt and
prompt leptons satisfying the loosened isolation requirement to also satisfy the tighter
one~\cite{ATLAS:2014ffa}. These efficiencies are measured in data 
control samples enriched in either fake/non-prompt leptons, mostly
multijet events, or prompt leptons, mostly $Z\to \ell^+\ell^-$ events. 
The shape of each multijet background distribution is derived by applying the same
method to the sample obtained with an identical selection as described
in Section~\ref{sec:analysis_strategy_overall}, but lowering the $b$-tagged-jet multiplicity 
requirement to two.
This strategy reduces the statistical uncertainty of the multijet background estimate, 
improving the stability of the fitting method described in 
Section~\ref{sec:stat_analysis}.


\section{Analysis strategy}
\label{sec:analysis_strategy_overall}

The $H \to 2a \to 4b$ decay chain is expected to have multiple $b$-tagged
jets, often three or four, satisfying the object selection. The
dominant background arises from \ttbar\ events. Preselected events 
are required to have exactly one electron or muon and at least three
jets, of which at least two must be $b$-tagged. 
Events are required to satisfy $\met>\unit[25]{\GeV}$ and the transverse mass\footnote{The transverse
mass is defined as $\mtw \equiv \sqrt{2 \met \pt^{\ell} (1-
  \cos{\Delta \phi})}$, where $\pt^{\ell}$ is the transverse
momentum of the lepton and $\Delta\phi$ is the azimuthal angle between
the lepton and $\ensuremath{\vec{E}_{\text{T}}^{\text{miss}}}$ directions.} must
fulfil $\mtw >\unit[50]{\GeV}$, in order to be consistent with $\Wboson$
boson decays. Events are categorised into
eight channels depending on the number of jets (3, 4 and $\geq$5) and the
number of $b$-tagged jets (2, 3 and $\geq$4). These analysis channels are
referred to as ($n$j, $m$b) indicating $n$ selected jets including $m$
$b$-tagged jets.

The categories most sensitive to the $H \to 2a \to 4b$ decay chain are 
(3j, 3b), (4j, 3b) and (4j, 4b). 
In these channels, background $\ttbar$ events can only satisfy the
selection criteria if accompanied by additional $b$-tagged
jets. In the case of (3j, 3b) or (4j, 3b), the main sources of
$\ttbar$ background are events with jets mis-identified as $b$-jets,
particularly from $W \to cs$ decays, where the $c$-jet is mis-identified, and from $W \to \tau\nu$, where the
$\tau$-lepton decays hadronically and is likewise mis-identified. 
In the case of (4j, 4b), the $\ttbar$ background includes
more events with genuine $b$-quarks from gluon splitting to $b\bar{b}$ pairs.
The main purpose of the five other jet and $b$-tagged-jet multiplicity
channels is to constrain the $\ttbar$+jets background prediction and the related systematic 
uncertainties (see Section~\ref{sec:systematics}) through a profile likelihood 
fit to data (see Section~\ref{sec:stat_analysis}). 

The $\ttbar$+light background is dominant 
in the sample of events with exactly two or three $b$-tagged jets. The background
processes $\ttcc$ and $\ttbb$ become more important as the jet and
$b$-tagged-jet multiplicities increase. In particular, the $\ttbb$ background
dominates for events with $\geq$5 jets and $\geq$4 $b$-tagged jets. 

\subsection{Signal and background discrimination}
\label{sec:analysis-strategy}

In order to improve the sensitivity of the search, several kinematic
variables are identified to distinguish between signal and
background, and are combined into a boosted decision tree 
(BDT) multivariate discriminant~\cite{scikit-learn} that uses the AdaBoost 
algorithm~\cite{Freund1997119}. The BDT is trained
to discriminate between signal events with an $a$-boson mass of
\unit[60]{\GeV} and $\ttbar$ events. As described below, the variables 
chosen as input for the BDT do not depend strongly on the value of
$m_a$ and provide excellent separation between signal and background, 
so training each mass hypothesis separately with these variables
would only slightly improve the sensitivity of the search. The training is
performed separately for each of the channels (3j, 3b), (4j, 3b) and
(4j, 4b) since the signal and background kinematics differ between them. 

Signal events are characterised by the
presence of a resonance resulting from the Higgs boson decay $H \to 2a
\to 4b$. Two variables are used to reconstruct particles from the signal
decay chain. The first is the reconstructed invariant mass of the
$b$-tagged jets, {$m_{bbb}$} or {$m_{bbbb}$}, defined for events with
three or four $b$-tagged jets respectively, which peaks around the
Higgs boson mass for signal events. In the case of three $b$-tagged
jets, the peak is due to events where two $b$-quarks are merged in a
single jet or one of the $b$-quarks is very soft in an asymmetric
decay and has a small impact on the kinematics. 
The second discriminating variable for events with four~$b$-tagged jets
is the minimum difference between the invariant masses of $bb$ pairs 
($\Delta m_{\rm min}^{bb}$). For signal events, two pairs of
$b$-quarks originate from a pair of $a$-bosons, so for the correct
jet pairing, $m_{bb} \approx m_a$, and the difference between the invariant 
masses of the $bb$ pairs is smaller for signal than for $\ttbar$
background events. 

Additional kinematic variables exhibit differences between signal and
background. The $H_{\rm T}$ variable, defined as the scalar sum of $\pt$ 
for all jets in the event, is a measure of the total hadronic energy in
the event, which is typically larger for $\ttbar$ than for $WH$ events. 
The transverse momentum of the $W$ boson, $\pt^W$, 
constructed from the vector sum of the $\ensuremath{\vec{E}_{\text{T}}^{\text{miss}}}$ and the
lepton $\ensuremath{\vec{p}_{\text{T}}}\xspace$, is slightly higher for signal $WH$ events, where the $W$ boson 
recoils against the Higgs boson, than for background \ttbar events.
Another variable used is the average angular separation between all
pairs of $b$-tagged jets,
referred to as $\Delta R_{\rm av}^{bb}$. For background $\ttbar$
events, the $b$-tagged jets originate from the decays of the two top
quarks and tend to be spatially more separated than for the
signal. A related variable is the minimum $\Delta R$ separation
between any $b$-tagged jet and the lepton, $\Delta R_{\rm min}^{\ell b}$. 
In $\ttbar$ background events, the lepton is typically closer to a $b$-tagged jet 
than in signal events, since the lepton and the nearest $b$-tagged jet often originate 
from the same top-quark decay.
In the case of the signal, the Higgs boson and hence the $b$-jets
recoil against the $W$ boson, which the lepton comes from. 

Finally, two variables are used to identify particles from the dominant $\ttbar$
background decay chain. The first variable is used in the (4j, 3b)
channel to distinguish between $\ttbar$ events with two $b$-tagged
jets from the top-quark decays and $\ttbar$ events with a third
$b$-tagged jet from a mis-identified charm or light jet from the
hadronically decaying $\Wboson$ boson. 
The invariant mass of two $b$-tagged jets, selected as the pair
with the smallest $\Delta R$ separation, and the non-$b$-tagged jet,
$m_{bbj}$, reconstructs the hadronically decaying top quark, peaking
around the top-quark mass for these background events. 
The second variable, used in the (4j, 4b) channel, is a variant of the $m_{\rm T2}$
observable, defined as the minimum ``mother'' particle mass compatible 
with all the transverse momenta and mass-shell
constraints~\cite{Burns:2008va}, that identifies events with several
invisible particles. In the case of the \ttbar background events, in
addition to the \met from the neutrino from a leptonic $W$ boson
decay, invisible particles may arise from a $\tau$-lepton decay or from a
lost jet from a $W$ boson. In these cases, the $m_{\rm T2}$ has an
endpoint at the top-quark mass, which is not the case for the signal.  

\begin{figure*}[htbp]
\begin{center}
\subfloat[]{\includegraphics[width=0.45\textwidth]{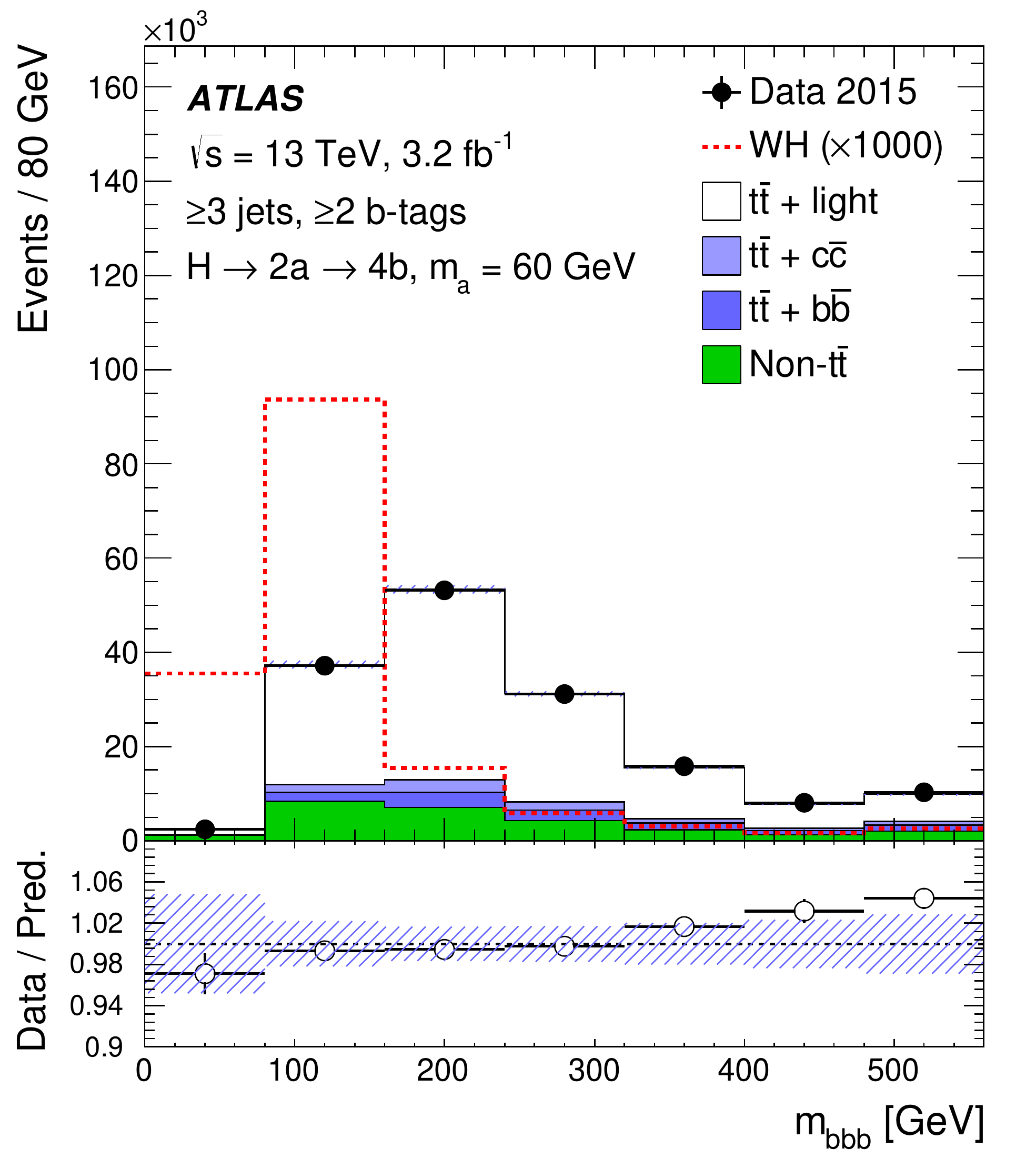}}%
\subfloat[]{\includegraphics[width=0.45\textwidth]{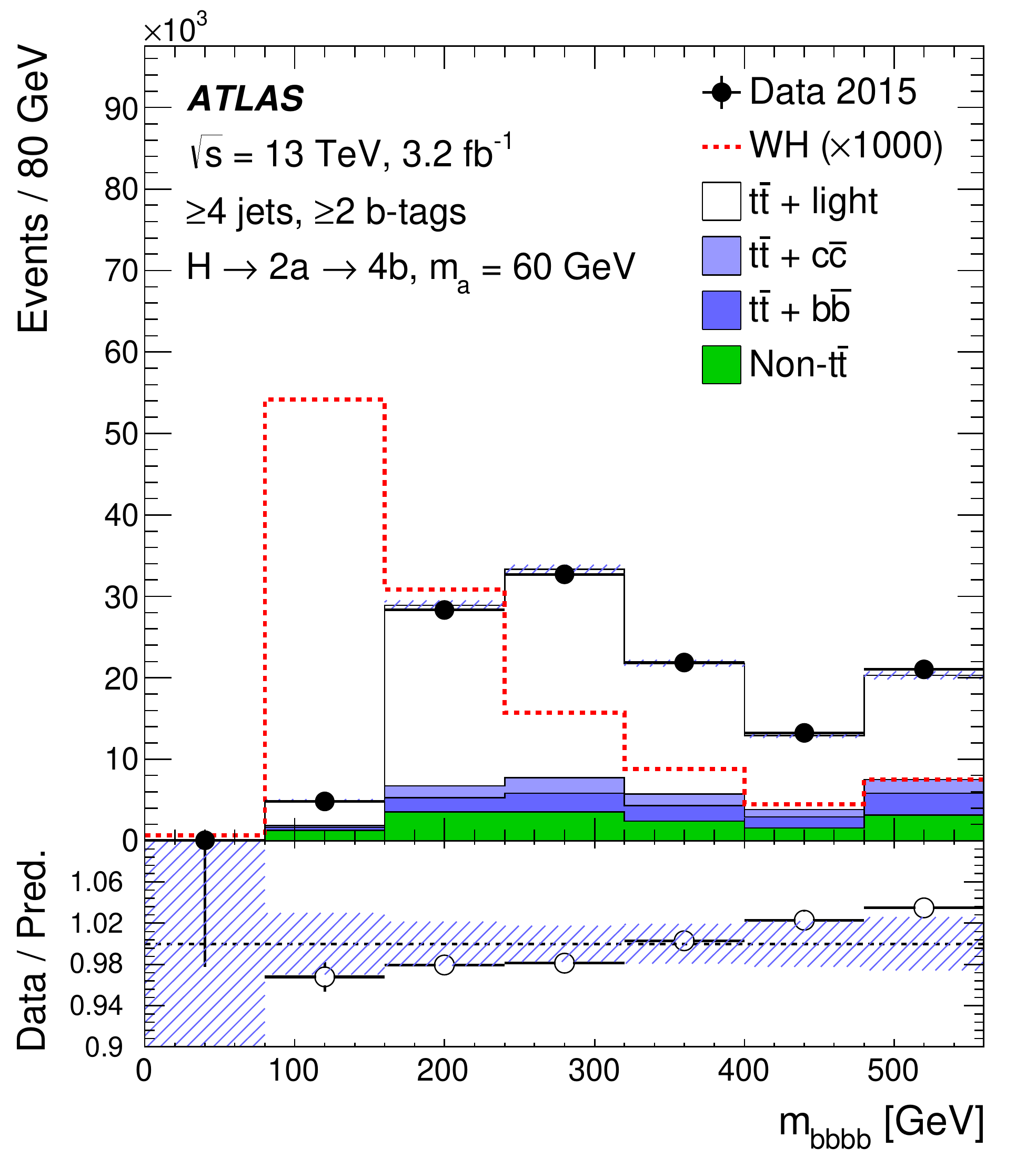}}\\%
\subfloat[]{\includegraphics[width=0.45\textwidth]{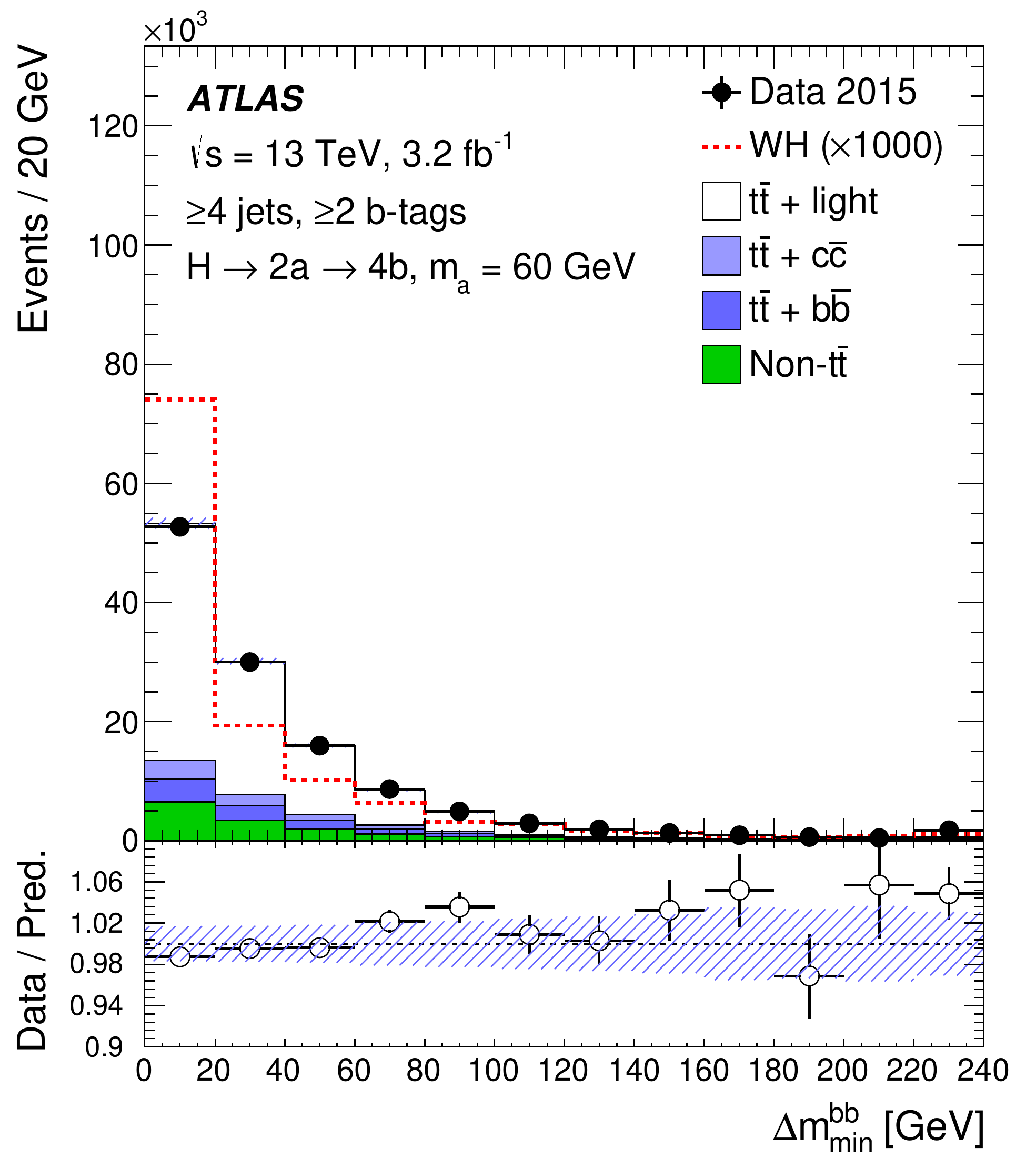}}
\caption{Comparison of data with the SM background predictions for 
the distributions of (a) $m_{bbb}$, (b) $m_{bbbb}$ and (c)
$\Delta m_{\rm min}^{bb}$ in the sample that is inclusive in 
number of jets and $b$-tagged jets. Distributions for the signal model
($WH$, $H\rightarrow 2a \rightarrow 4b$), with $m_a=$ 60 GeV, normalised to the SM $pp
\rightarrow WH$ cross section, assuming BR$(H\rightarrow aa)$ $\times$ BR$(a
\rightarrow bb)^2 = 1$ and scaled by a factor of 1000, are overlaid. 
The hashed area represents the total uncertainty in the
background. Comparisons use events with 
$\ge3$ jets, except when at least four jets are necessary to define
the variable, in which case events with $\ge4$ jets are used. The last
bin contains the overflow. Markers are not drawn if they are outside the y-axis range.}
\label{fig:KinVars_preselection1}
\end{center}
\end{figure*}

\begin{figure*}[htbp]
\begin{center}
\subfloat[]{\includegraphics[width=0.45\textwidth]{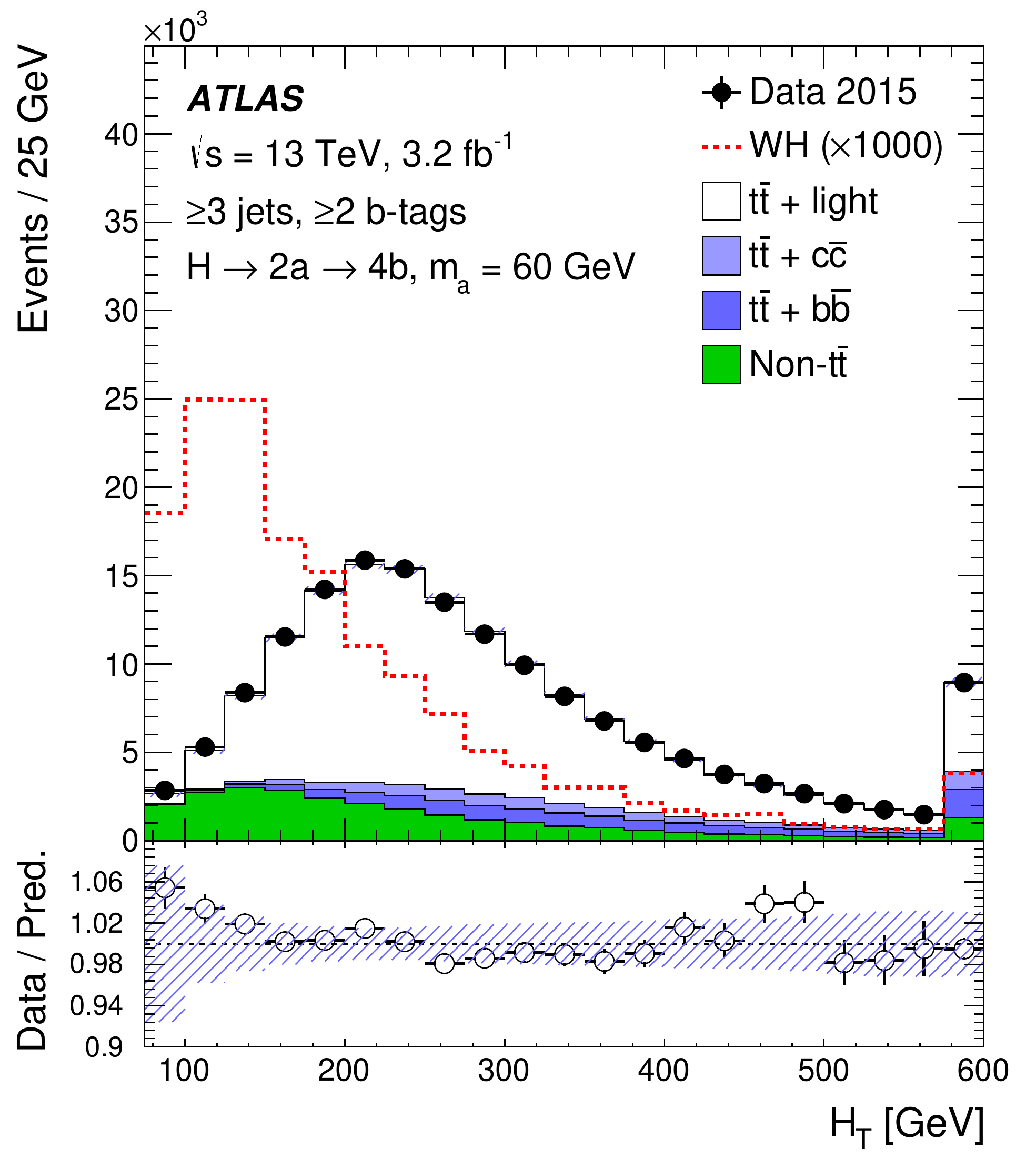}}%
\subfloat[]{\includegraphics[width=0.45\textwidth]{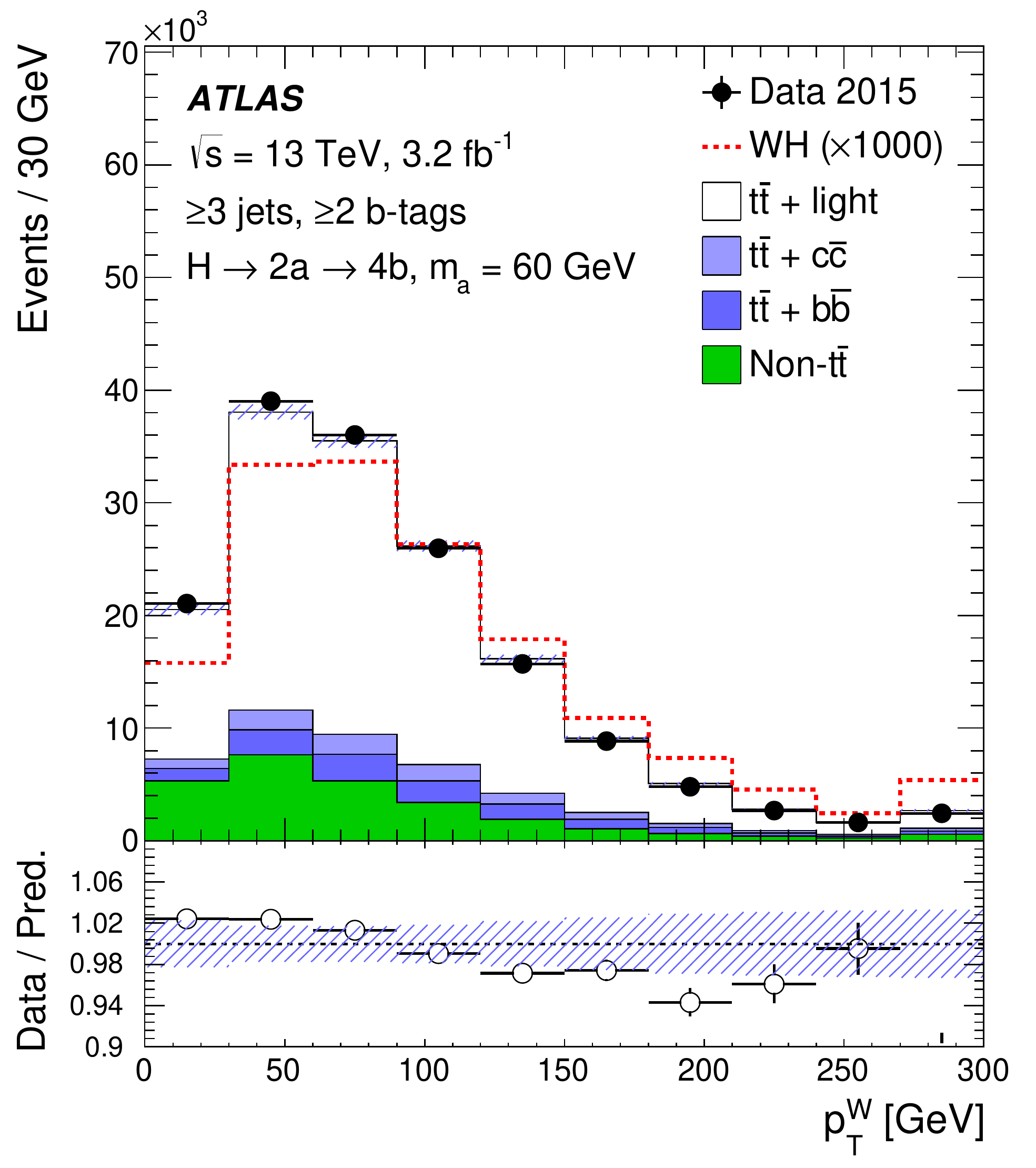}}\\%
\subfloat[]{\includegraphics[width=0.45\textwidth]{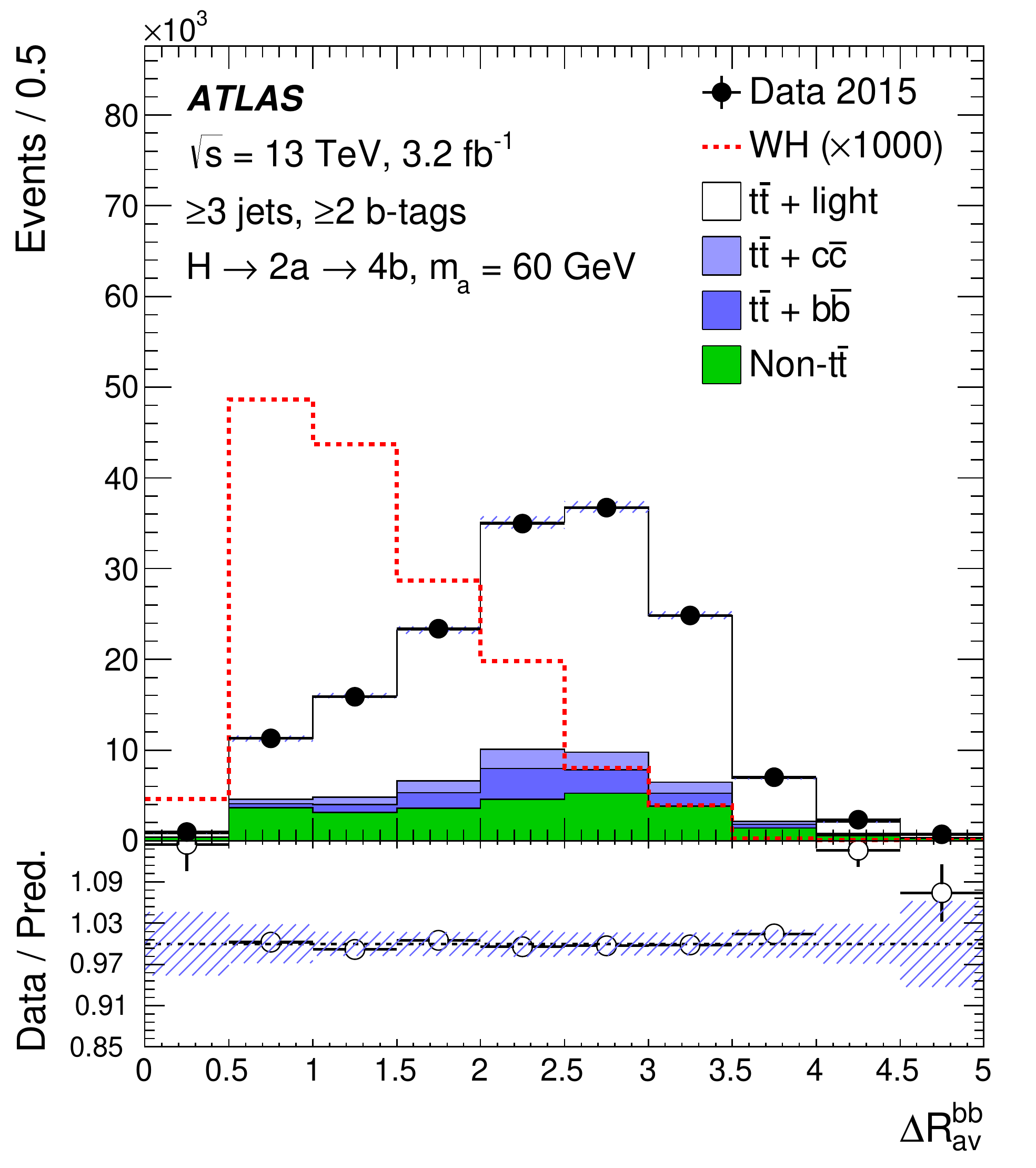}}
\subfloat[]{\includegraphics[width=0.45\textwidth]{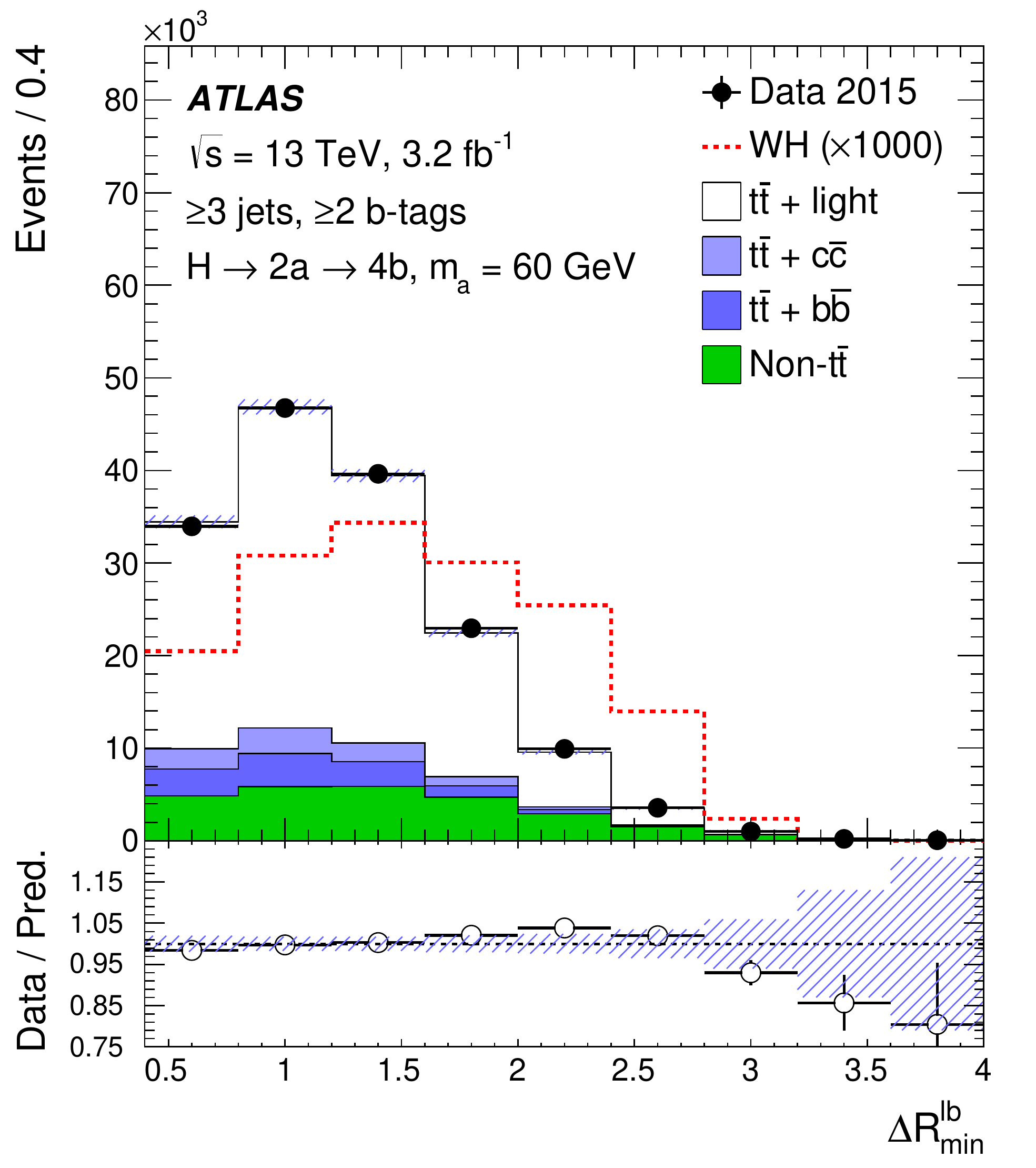}}%
\caption{Comparison of data with the SM background predictions for 
the distributions of (a) $H_{\rm T}$, (b) $\pt^W$, 
(c) $\Delta R_{\rm av}^{bb}$ and (d) $\Delta R_{\rm min}^{\ell b}$ in the sample that is inclusive in 
number of jets and $b$-tagged jets. Distributions for the signal model ($WH$, $H\to 2a \to
4b$), with $m_a=\unit[60]{\GeV}$, normalised to the SM $pp
\rightarrow WH$ cross section, assuming $\BR(H\to aa) \times \BR(a
\to bb)^2 = 1$ and scaled by a factor of 1000, are overlaid. 
The hashed area represents the total uncertainty in the
background. The last bin contains the overflow.
} 
\label{fig:KinVars_preselection2}
\end{center}
\end{figure*}

\begin{figure*}[htbp]
\begin{center}
\subfloat[]{\includegraphics[width=0.45\textwidth]{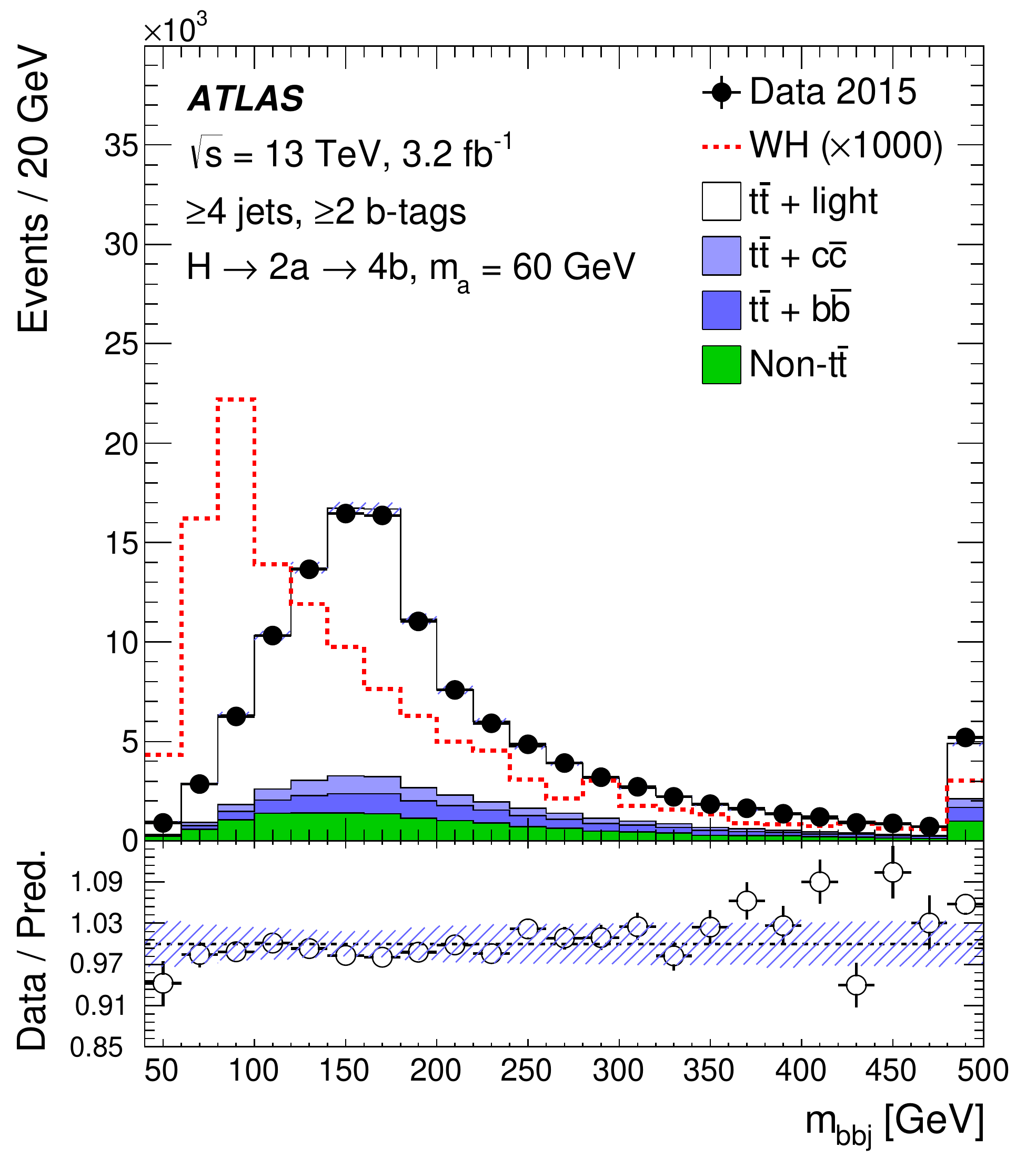}}%
\subfloat[]{\includegraphics[width=0.45\textwidth]{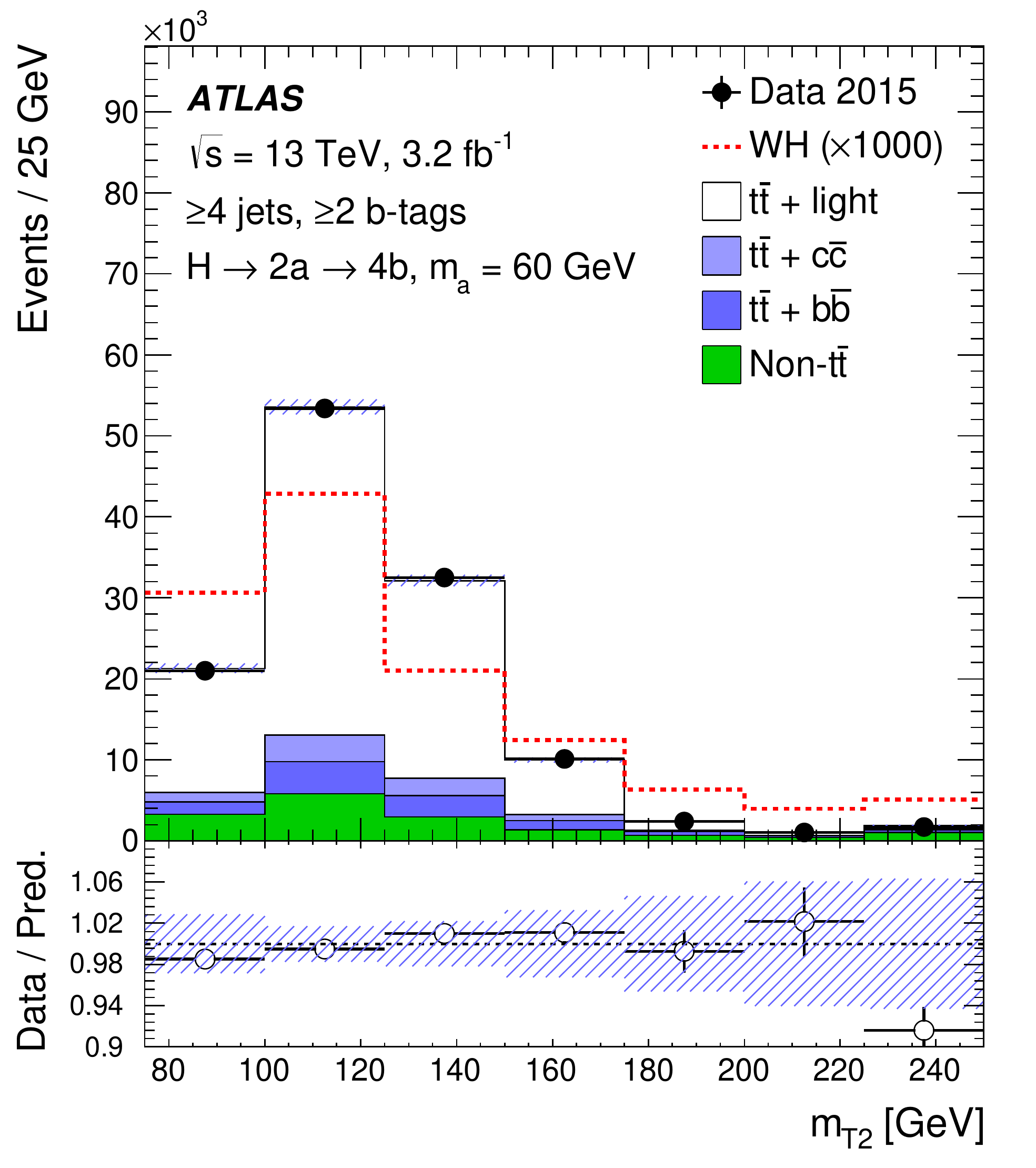}}
\caption{Comparison of data with the SM background predictions for 
the distributions of (a) $m_{bbj}$ and (b) $m_{\rm T2}$ in the sample that is inclusive in 
number of jets and $b$-tagged jets. Distributions for the signal model ($WH$, $H\to 2a \to
4b$), with $m_a=\unit[60]{\GeV}$, normalised to the SM $pp
\rightarrow WH$ cross section, assuming $\BR(H\to aa) \times \BR(a
\to bb)^2 = 1$ and scaled by a factor of 1000, are overlaid. 
The hashed area represents the total uncertainty in the
background. The last bin contains the overflow.
} 
\label{fig:KinVars_preselection3}
\end{center}
\end{figure*}

\begin{figure*}[htbp]
\begin{center}
\subfloat[]{\includegraphics[width=0.45\textwidth]{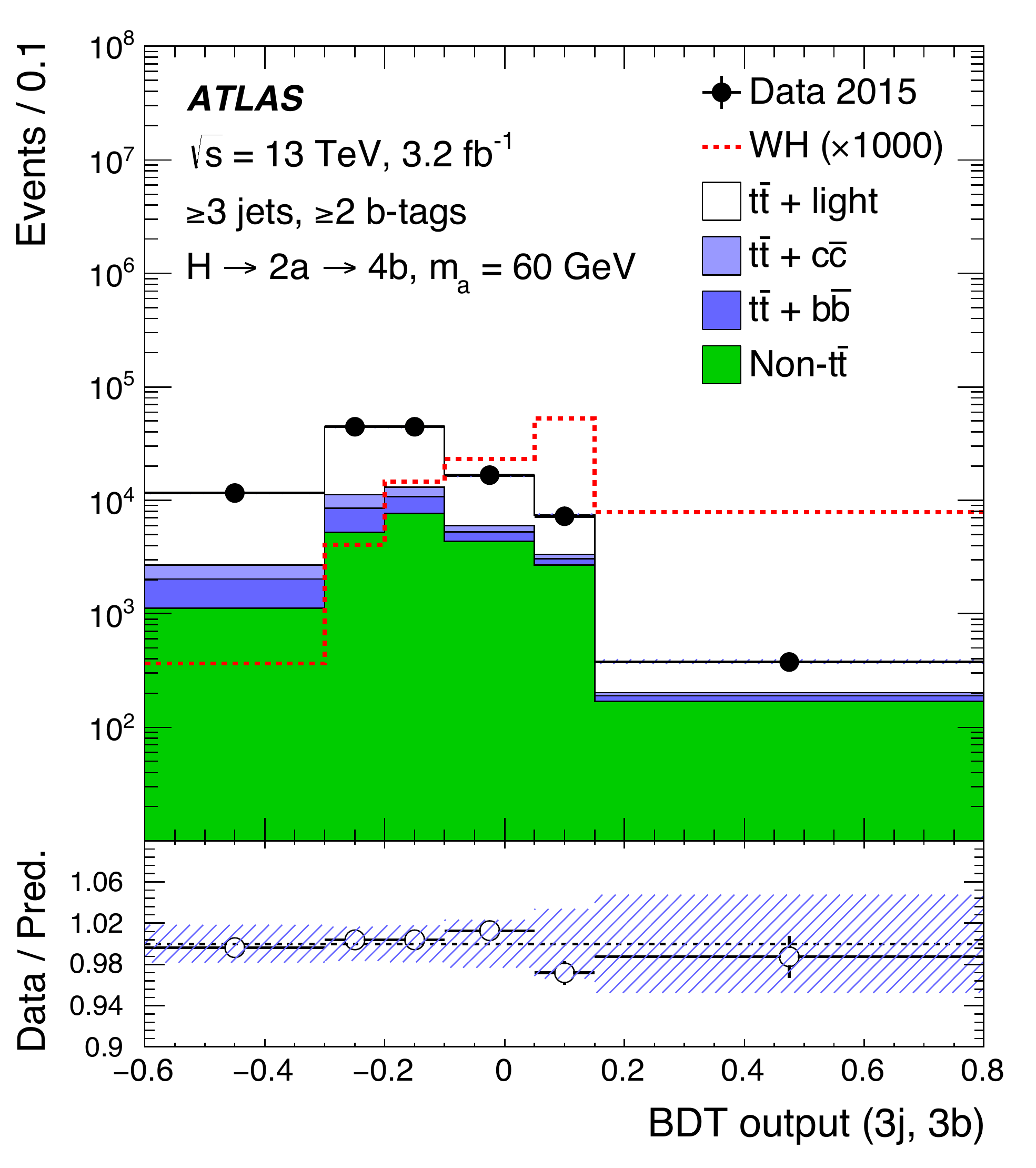}}%
\subfloat[]{\includegraphics[width=0.45\textwidth]{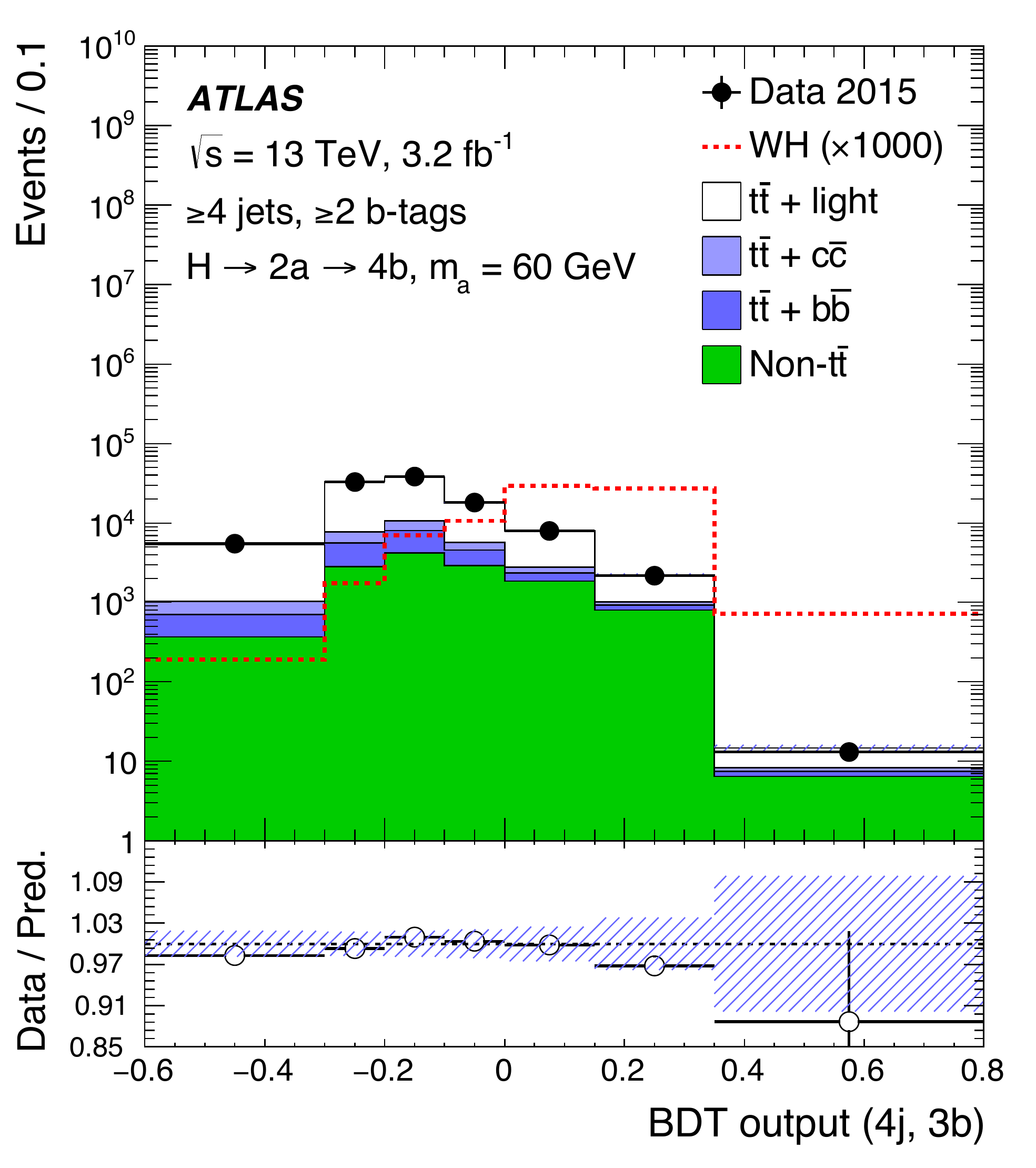}}\\%
\subfloat[]{\includegraphics[width=0.45\textwidth]{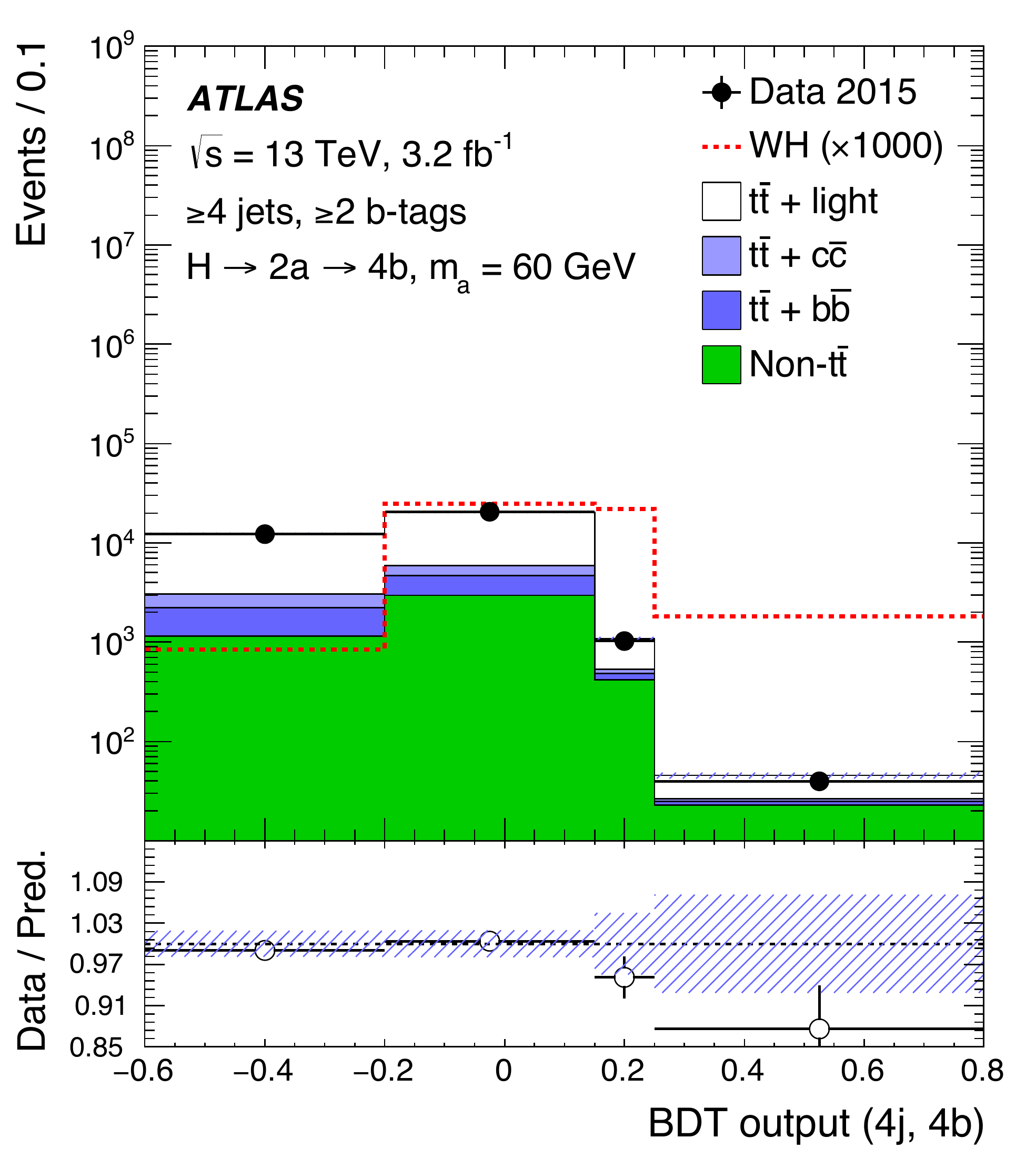}}
\caption{Comparison of data with the SM background predictions for 
the distributions of (a) BDT (3j, 3b), (b) BDT (4j, 3b), and (c) BDT (4j, 4b) in the sample that is inclusive in 
number of jets and $b$-tagged jets. Distributions for the signal model
($WH$, $H\rightarrow 2a \rightarrow 4b$), with $m_a=$ 60 GeV, normalised to the SM $pp
\rightarrow WH$ cross section, assuming BR$(H\rightarrow aa)$ $\times$ BR$(a
\rightarrow bb)^2 = 1$ and scaled by a factor of 1000, are overlaid. 
The hashed area represents the total uncertainty in the
background. Comparisons use events with 
$\ge3$ jets, except when at least four jets are necessary to define
the BDT discriminant, in which case events with $\ge4$ jets are
used. The BDT output is determined in the range $[-1,1]$. The first 
and last bin contain the underflow and overflow, respectively.}
\label{fig:BDT_preselection}
\end{center}
\end{figure*}

Table~\ref{tab:BDTVars} indicates which variables are used to train each of the
three BDT discriminants for the (3j, 3b), (4j, 3b), and (4j, 4b)
categories. Figures~\ref{fig:KinVars_preselection1}--\ref{fig:KinVars_preselection3} show the expected
distributions of the kinematical variables obtained after using the statistical procedure
and the systematic uncertainties described in Sections~\ref{sec:stat_analysis} 
and~\ref{sec:systematics}, respectively.
These variables are used in the BDT discriminants for signal and
background for all events that satisfy the event selection criteria,
and are shown in
Figures~\ref{fig:KinVars_preselection1}--\ref{fig:KinVars_preselection3}
inclusively in number of jets and $b$-tagged jets. The distributions
are dominated by events with the minimum number of
$b$-tagged jets. In this comparison, the jets in each event are
ordered by value of the $b$-tagging discriminant and those with the 
highest score are used to calculate the input variables of the BDT,
even if they do not satisfy the $b$-tagging criteria used in this
analysis. The distributions are similar to those obtained in each 
analysis channel and indicate that each variable individually has 
some signal and background discrimination power. The tail in the 
$m_{bbbb}$ distribution for signal events, shown in
Figure~\ref{fig:KinVars_preselection1}, is mainly formed by events
with jets mis-associated to the $a$-boson decay. The tail is greatly reduced
in the signal regions with the tighter requirement on the number of
$b$-tagged jets. Figure~\ref{fig:BDT_preselection}
shows the BDT discriminant for signal and background events that
satisfy the event selection criteria inclusively in number of jets and
$b$-tagged jets. These distributions are used to validate the BDT
modelling in background-enriched samples with kinematic properties
that are similar to those in the signal regions. 

\begin{table}[htbp]
\begin{center}
\begin{small}
  \begin{tabular*}{\textwidth}{@{\extracolsep{\fill}}ll|ccccccccc}
\toprule
\multicolumn{2}{l|}{Region} &  $m_{bbb}$   & $m_{bbbb}$  & $\Delta m_{\rm min}^{bb}$& $H_{\rm T}$ & $\pt^W$ & $\Delta R_{\rm av}^{bb}$ & $\Delta R_{\rm min}^{\ell b}$ &  $m_{bbj}$ & $m_{\rm T2}$ \\
\hline
\multirow{3}{*}{\phantom{0}Signal} & (3j, 3b) & \checkmark &                    & & \checkmark & \checkmark & \checkmark & \checkmark & & \\
&(4j, 3b) & \checkmark &                    & & \checkmark & \checkmark & \checkmark& & \checkmark& \\
&(4j, 4b) & & \checkmark &  \checkmark &\checkmark& & \checkmark &  & & \checkmark\\
\hline
\multicolumn{2}{l|}{Control} & & & & \checkmark & & & & & \\ 
\bottomrule
\end{tabular*}
\caption{\label{tab:BDTVars} List of variables used in the three signal
  regions as inputs to the BDT multivariate discriminant and used
  in the five control regions. The variables are described in the text.}
  \end{small}
  \end{center}
\end{table}

\subsection{Fitting procedure}
\label{sec:stat_analysis}

The distributions of the final discriminants in the eight analysis channels 
considered are combined to test the presence of a signal. The BDT
discriminant, described in Section~\ref{sec:analysis-strategy}, is used
for the channels enriched with signal, (3j, 3b),
(4j, 3b) and (4j, 4b), while the $\HT$ distribution is used in the five control channels. 
The statistical analysis is based on a binned
likelihood function constructed as a product of
Poisson probability terms over all bins considered in the search. 

The likelihood function, $L$, depends on the parameter of
interest, the signal-strength $\mu$, defined as:
\begin{equation}
\mu = \sigma(WH)\times \BR(H\to aa) \times \BR(a \to bb)^2,
\label{eq:br_mu}
\end{equation}
where $\sigma(WH)$ is the production cross section for $pp \to WH$.

Systematic uncertainties in the signal and background predictions 
(see Section~\ref{sec:systematics}) are accounted for in the likelihood function as a set of nuisance
parameters, $\bm{\theta}$. These parameters are implemented as Gaussian 
priors in the case of shape uncertainties and log-normal priors for 
uncertainties affecting the normalisation, with width parameters
corresponding to the size of the respective uncertainties. Statistical uncertainties in the background estimates in each bin of
the discriminant distributions are also taken into account via
dedicated nuisance parameters in the fit.     

The background-only hypothesis is tested by fitting the background
predictions to the observed data, setting $\mu=0$ 
and maximising the likelihood over $\bm{\theta}$.
The best-fit $\mu$ is obtained by performing a binned likelihood fit to the data under the signal-plus-background
hypothesis, i.e. maximising the likelihood function $L(\mu,\bm{\theta})$
over $\mu$ and $\bm{\theta}$.
The nuisance parameters $\bm{\theta}$ allow variations of the predicted signal and background
according to the corresponding systematic uncertainties, and their 
fitted values correspond to the deviations from the nominal
predictions that globally provide the best fit to the data.
This procedure allows a reduction of the impact of systematic 
uncertainties on the search sensitivity by taking advantage of the 
highly populated background-dominated channels included in the likelihood fit.


\section{Systematic uncertainties}
\label{sec:systematics}

Several sources of systematic uncertainty are considered that affect
the normalisation or the shape of the signal and background
contributions to the final discriminant distributions. 
Each source of systematic uncertainty is considered to be uncorrelated
with other sources, but correlated across processes and channels 
where appropriate. This section describes the 
sources of systematic uncertainty considered in this search.

{\bf Luminosity and pile-up:} The uncertainty in the integrated
luminosity is $5\%$, affecting the overall normalisation of
all processes estimated from the simulation. 
It is derived, following a methodology similar to
that detailed in Ref.~\cite{Aad:2013ucp}, from a calibration of the
luminosity scale using $x$--$y$ beam-separation scans performed
in August 2015. 
The uncertainty associated with the modelling of pile-up 
arises mainly from differences between the expected and observed 
fraction of the visible $pp$ cross section.

{\bf Reconstructed objects:} Uncertainties associated with leptons arise from the reconstruction,
identification and trigger efficiencies, as well as lepton momentum
scales and resolutions. 
These efficiencies are measured using tag-and-probe techniques on 
$Z\to \ell^+\ell^-$ data and simulated events. 
The small differences found are corrected in the simulation.
Negligible uncertainties arise from the corrections
applied to adjust the lepton momentum scales and resolutions in 
simulation to match those in data. 
The combined effect of all these uncertainties results in an overall normalisation 
uncertainty in the signal and background of less than $1\%$.

Uncertainties associated with jets arise from the efficiency of jet reconstruction 
and identification, as well as the jet energy scale
and resolution. The largest contribution comes from the jet energy
scale uncertainty, which depends on jet $\pt$ and $\eta$. 
It affects the normalisation of signal and backgrounds by approximately $5\%$
in the most sensitive search channels.
Uncertainties associated with energy scales and resolutions of leptons and jets 
are propagated to $\met$. An uncertainty in the contribution from
charged-particle tracks is also included in the $\met$
uncertainty~\cite{ATL-PHYS-PUB-2015-027}. Additional uncertainties originating from the modelling 
of the underlying event are negligibly small.

Several uncertainties are associated with the identification of the jet flavour, in
particular the modelling of the $b$-, $c$-, and light-jet-tagging
efficiencies in the simulation, which are corrected to match the
efficiencies measured in data~\cite{Aad:2015ydr,btag-Run2,ATL-PHYS-PUB-2015-022}.
These uncertainties are derived from studies performed with data at $\sqrt{s} = \unit[8]{\TeV}$
and are extrapolated to $\unit[13]{\TeV}$. They depend on the jet $\pt$ and the light-jet-tagging
additionally depends on the jet $\eta$. The sources of systematic uncertainty in
the tagging efficiencies are taken as uncorrelated between $b$-jets, $c$-jets, and
light-jets. They have their largest impact in the (4j, 4b) channel,
resulting in $4\%$ uncertainty in the $\ttbb$ background normalisation 
associated with the uncertainty in the $b$-jet-tagging scale factors, $8\%$
uncertainty in the $\ttcc$ background normalisation associated with
the uncertainty in the $c$-jet-tagging scale factors, and $45\%$ uncertainty
in the normalisation of the $\ttbar$+light background normalisation
associated with the uncertainty in the light-jet-tagging scale factors.

{\bf Background modelling:} Several sources of systematic
uncertainty affecting the modelling of $\ttbar$+jets are considered. 
An uncertainty of approximately 6\% is assumed for the $\ttbar$ production
cross section~\cite{Czakon:2011xx}, including contributions from
variations of the factorisation and renormalisation scales, and
uncertainties arising from the PDFs, $\alpha_{\rm S}$, and the
top-quark mass.

A 50\% uncertainty is assigned to the normalisation of the $\ttbb$
background. This uncertainty is derived from a comparison of the $\ttbb$
production cross sections predicted by {\powheg}+{\pythia} and by {\sc
  Sherpa+OpenLoops} at NLO (see Section~\ref{sec:bkg_modeling})~\cite{Aad:2015gra}. An additional 
$50\%$ uncertainty is assigned to the component of the $\ttbb$
background that contains exactly one $b$-hadron not originating from a top-quark
decay matched to a particle jet. The same systematic uncertainty of 
$50\%$ is applied to the normalisation of the $\ttcc$ background in the absence of an NLO
prediction for this process. The uncertainties in the $\ttbb$ components 
and $\ttcc$ are treated as uncorrelated. 

Systematic uncertainties affecting the shape of the $\ttbar$
background account for the choice of generator, the choice of parton
shower and hadronisation models, and the effects of initial- and
final-state radiation. The uncertainties are derived from comparisons between the 
nominal simulation and alternative samples produced with {\powheg} or 
{\amcatnlo} interfaced to {\pythia} or {\herwigpp} (see 
Section~\ref{sec:bkg_modeling}) and are treated as uncorrelated
across $\ttbar$+jets backgrounds. Additional uncertainties are
evaluated to account for the use of {\sherpa}+{\sc OpenLoops} NLO to 
model the $\ttbb$ background. In
particular, uncertainties are assessed for the PDFs, as well as the
choice of shower recoil
model and scale. An additional uncertainty accounts for limited 
knowledge of the component of the $\ttbb$ background originating 
from multiple parton interactions, which is not included in the NLO 
prediction. These systematic uncertainties are estimated 
following the methods described in Ref.~\cite{Aad:2015gra}.

The uncertainties in the predictions for the total cross sections
for the other background processes are applied
as normalisation uncertainties and are: $5\%$ for each of the
$W/Z$+jets and diboson processes, $+5\%$/$-4\%$ for single-top-quark production, $15\%$
for $\ttbar +\gamma/\Wboson/\Zboson$ and $+9\%$/$-12\%$ for $\ttbar
H$~\cite{Campbell:1999ah,Kidonakis:2011wy,Kidonakis:2010ux,Kidonakis:2010tc,Campbell:2012dh,Garzelli:2012bn,Dittmaier:2011ti,ATL-PHYS-PUB-2016-005}.
An additional uncertainty of $24\%$ is added in quadrature for
each additional jet to account for the extrapolation to higher jet
multiplicities, based on a comparison among different algorithms for
merging LO matrix-element and parton shower simulations~\cite{Alwall:2007fs}.
An uncertainty is applied to the modelling of the single-top-quark
background to account for the choice of scheme to handle the
overlaps between the \ttbar\ and $Wt$ final states. 
Small uncertainties arising from scale variations, which
change the amount of initial-state radiation and thus the event
kinematics, are also considered.

Uncertainties in the estimate of the multijet background come from the
limited number of events in the data sample without the isolation
requirement and from uncertainties in the measured non-prompt and
prompt lepton efficiencies. The normalisation uncertainty assigned to this
background is 60\%, as derived by comparing the multijet background 
prediction to data in control regions obtained by inverting the
requirements on the $\met$ and on $\mtw$. An uncertainty in the shape
of the predicted background distribution covers the difference between
the prediction obtained by reducing the required number of $b$-tagged jets and
the prediction at high $b$-tagged-jet multiplicity (see Section~\ref{sec:bkg_modeling}).

{\bf Signal modelling:} Several sources of systematic
uncertainty affect the theoretical modelling of the signal
acceptance. Uncertainties originate from the choice of PDFs, the factorization
and renormalization scales, and the parton shower, hadronisation and
underlying event models. 

As described in Section~\ref{sec:stat_analysis}, a binned maximum-likelihood fit is performed on the distributions of the final
discriminant in the eight channels considered. 
The fit constrains systematic uncertainties from several sources thanks to
the large number of events in the analysis channels considered and the
variations in the background composition across channels. The channels 
with two $b$-tagged jets constrain the main uncertainties 
affecting the $\ttbar$+light background prediction, while the
channels with $\geq$5 jets and $\geq$3 $b$-tagged jets are sensitive 
to the dominant uncertainties affecting the $\ttbar$+HF background 
prediction. 

After performing the fit, the leading sources of systematic
uncertainty are the modelling of the $\ttbar$+jets background and
$b$-, $c$- and light-jet-tagging efficiencies. Table~\ref{tab:SystSummary_H4b}
summarises the systematic uncertainties by indicating their impact on 
the normalisation of the signal and the main backgrounds in the (4j,
4b) channel. The uncertainties for the other signal channels (3j, 3b)
and (4j, 3b) are reduced to about $7\%$ for the $\ttbar+$light
contribution, mainly due to the reduced dependence on the light-jet-tagging
efficiency, and to about $12\%$ for the signal, primarily because of the
reduced $b$-tagging efficiency uncertainty due to the lower $b$-tagged-jet multiplicity requirement.

\begin{table}[htbp]
\begin{center}
\begin{small}
  \begin{tabular*}{\textwidth}{@{\extracolsep{\fill}}lcccc}
  \toprule
  Systematic uncertainty [\%] & $WH$, $H \to 2a \to 4b$ & $t\bar{t}$ + light & $t\bar{t} + c\bar{c}$ & $t\bar{t} + b\bar{b}$\\ 
\hline
  Luminosity & $4$ & $4$ & $4$ & $4$ \\
  Lepton efficiencies & $1$ & $1$ & $1$ & $1$ \\
  Jet efficiencies & $6$ & $4$ & $4$ & $4$ \\
  Jet energy resolution & $5$ & $1$ & $3$ & $1$ \\
  Jet energy scale & $4$ & $2$ & $4$ & $3$ \\
  $b$-tagging efficiency & $17$ & $5$ & $5$ & $9$ \\
  $c$-tagging efficiency & $1$ & $6$ & $12$ & $4$ \\
  Light-jet-tagging efficiency & $2$ & $29$ & $5$ & $3$ \\
  Theoretical cross sections & -- & $5$ & $5$ & $5$ \\
  $t\bar{t}$: modelling & -- & $6$ & $45$ & $26$ \\ 
  $t\bar{t}$+HF: normalisation & -- & -- & $35$ & $18$ \\
  $t\bar{t}$+HF: modelling & -- & -- & -- & $5$ \\
  Signal modelling & $7$ & -- & -- & -- \\
  \hline
  Total & $21$ & $31$ &  $54$ & $21$ \\
\bottomrule
  \end{tabular*}
  \caption{Summary of the impact of the considered systematic
    uncertainties (in \%) on the normalisation of the signal ($m_a$ = \unit[60]{\GeV}) and the 
    main backgrounds for the (4j, 4b) channel after the fit.
  The total uncertainty can differ from the sum in quadrature of individual sources due to correlations between them.}
  \label{tab:SystSummary_H4b}
  \end{small}
  \end{center}
\end{table}


\section{Results}
\label{sec:results}

\begin{figure*}[htbp]
\begin{center}
\subfloat[]{\includegraphics[width=0.45\textwidth]{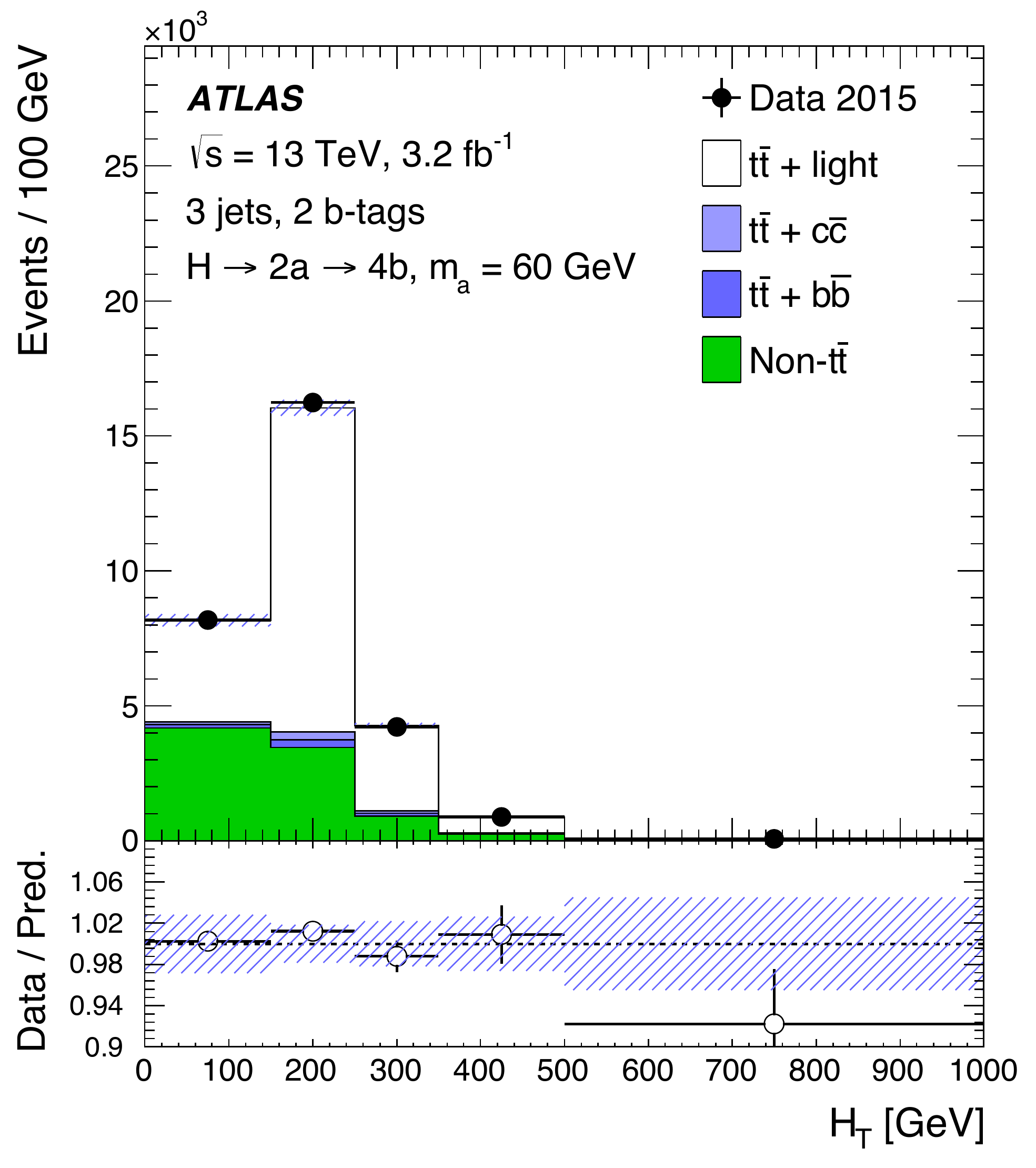}}%
\subfloat[]{\includegraphics[width=0.45\textwidth]{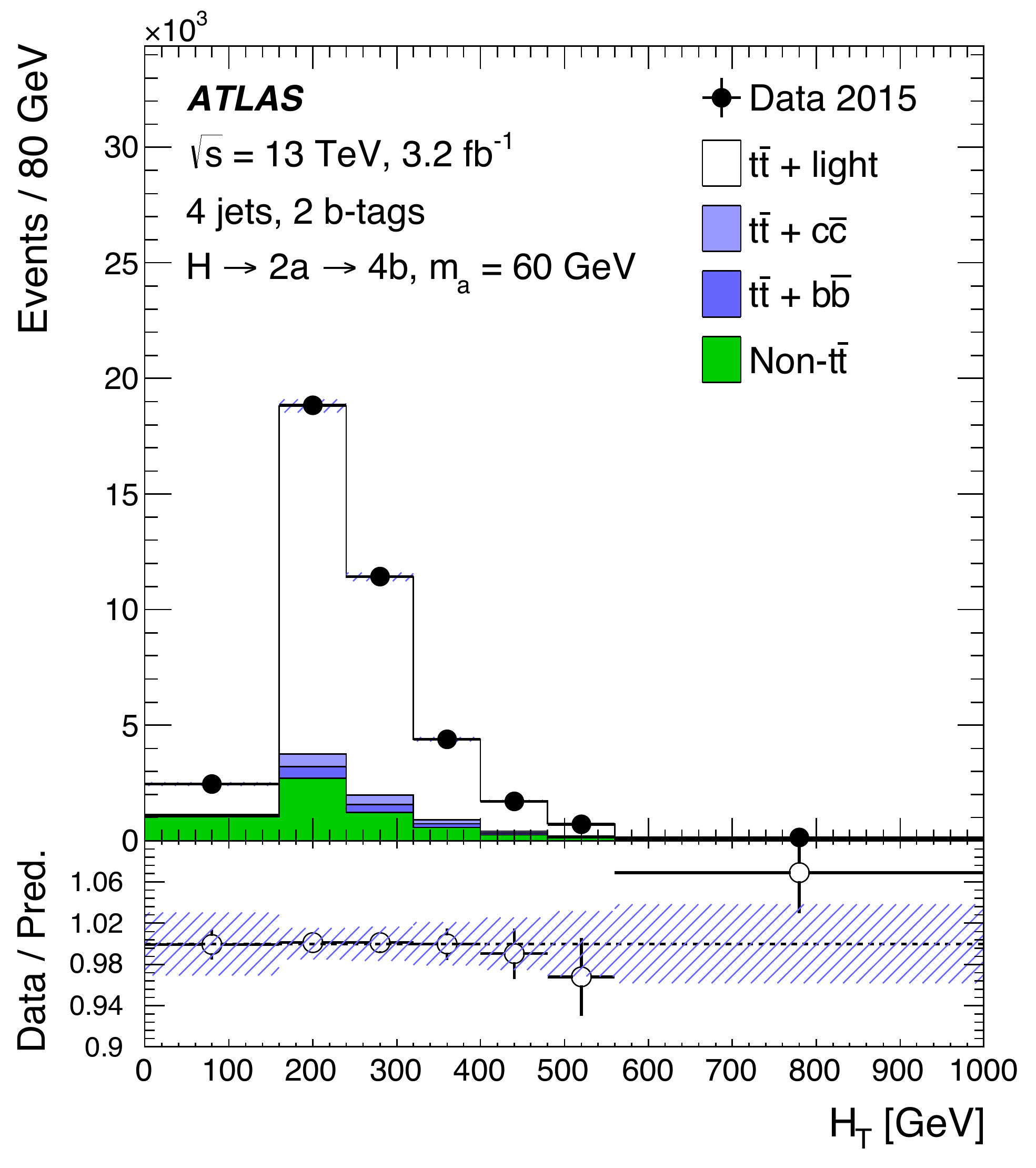}}\\%
\subfloat[]{\includegraphics[width=0.45\textwidth]{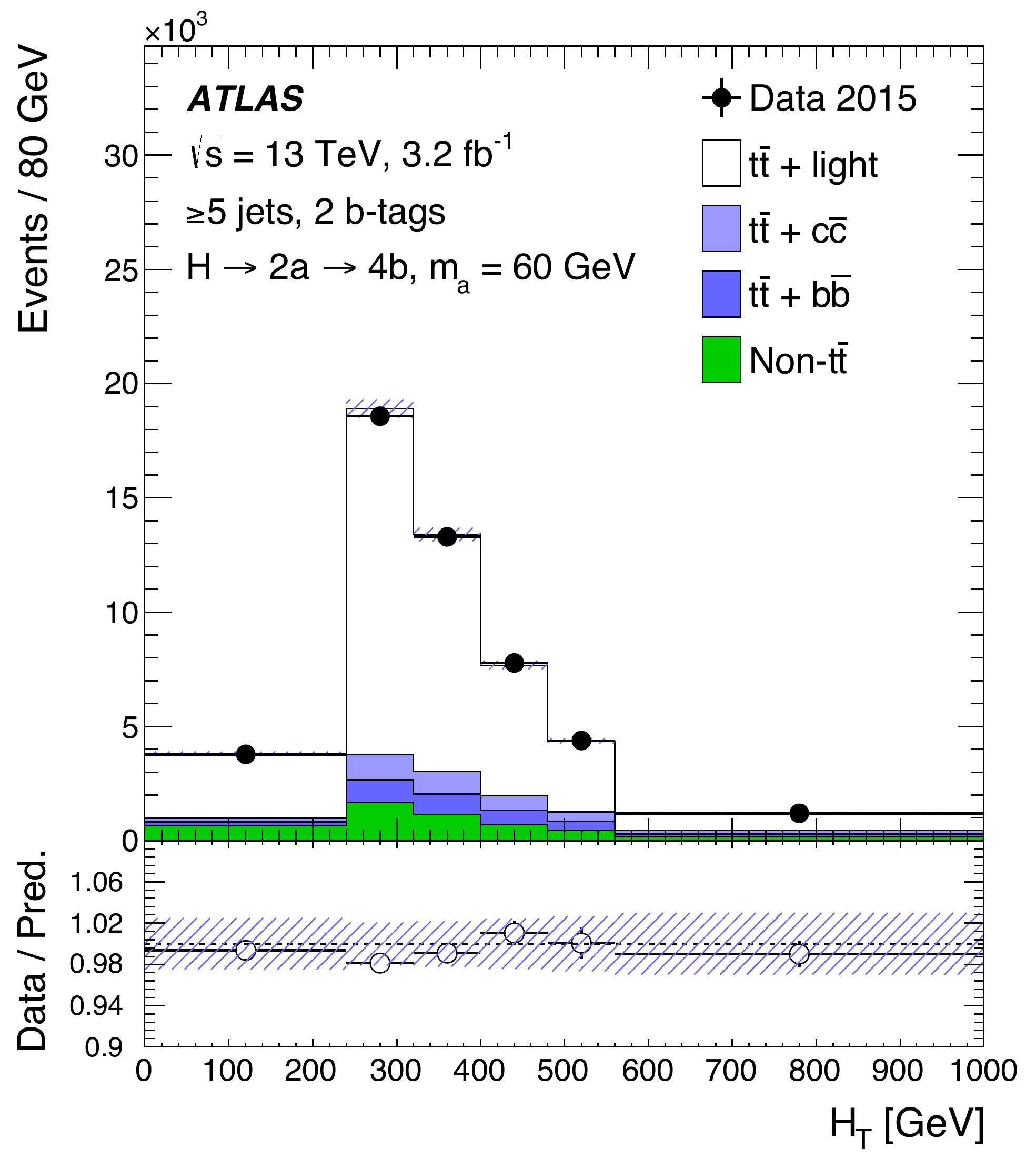}}
\caption{Comparison between the data and prediction for the
  distribution of the $H_{\rm T}$ variable used in the control regions
  with two $b$-tagged jets. 
  These distributions are after the fit is performed
  on data under the background-only hypothesis. The hashed area 
  represents the total uncertainty in the background. The last bin contains the overflow.
}   
\label{fig:postfit_unblinded1}
\end{center}
\end{figure*}

\begin{figure*}[htbp]
\begin{center}
\subfloat[]{\includegraphics[width=0.45\textwidth]{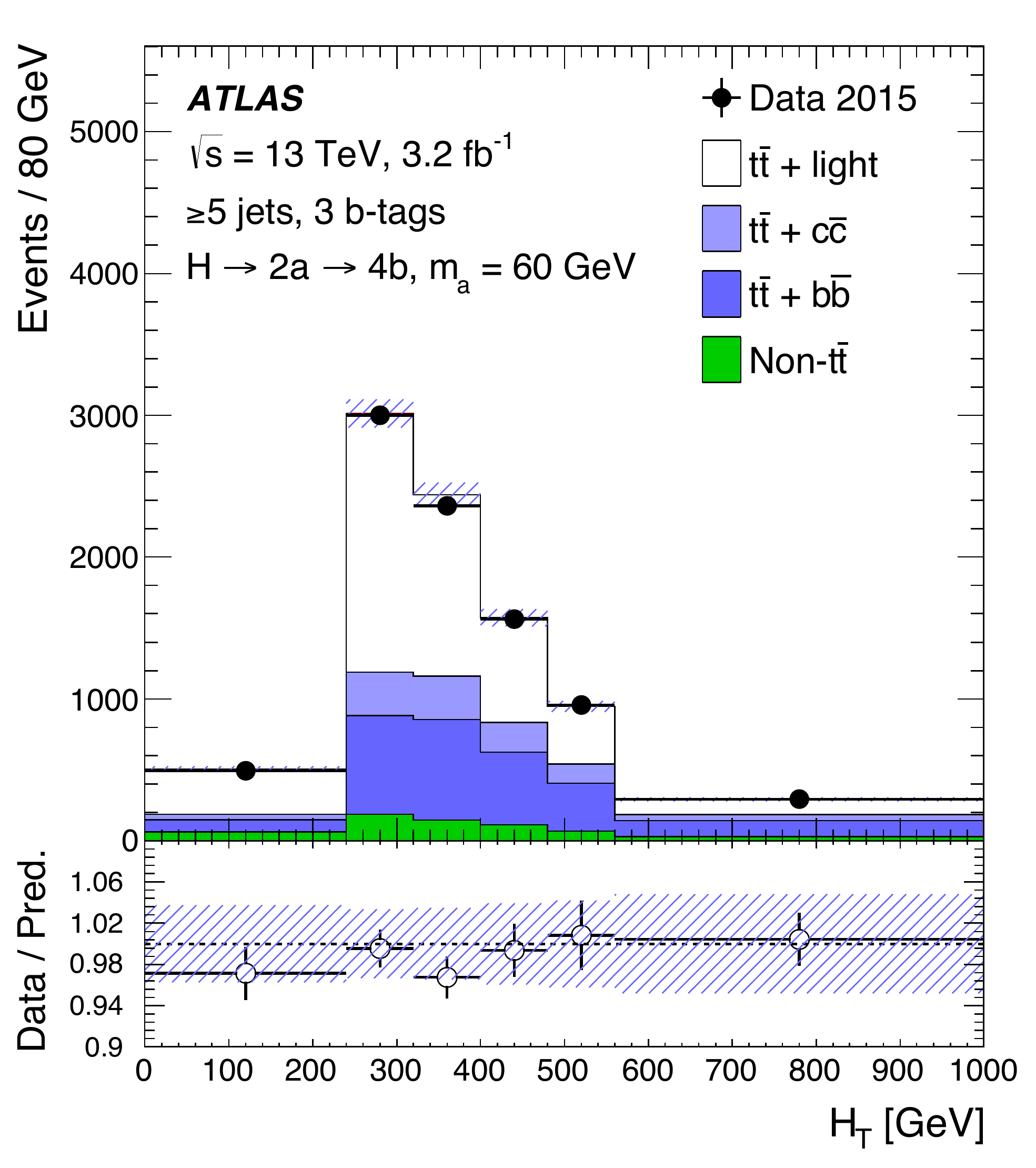}}%
\subfloat[]{\includegraphics[width=0.45\textwidth]{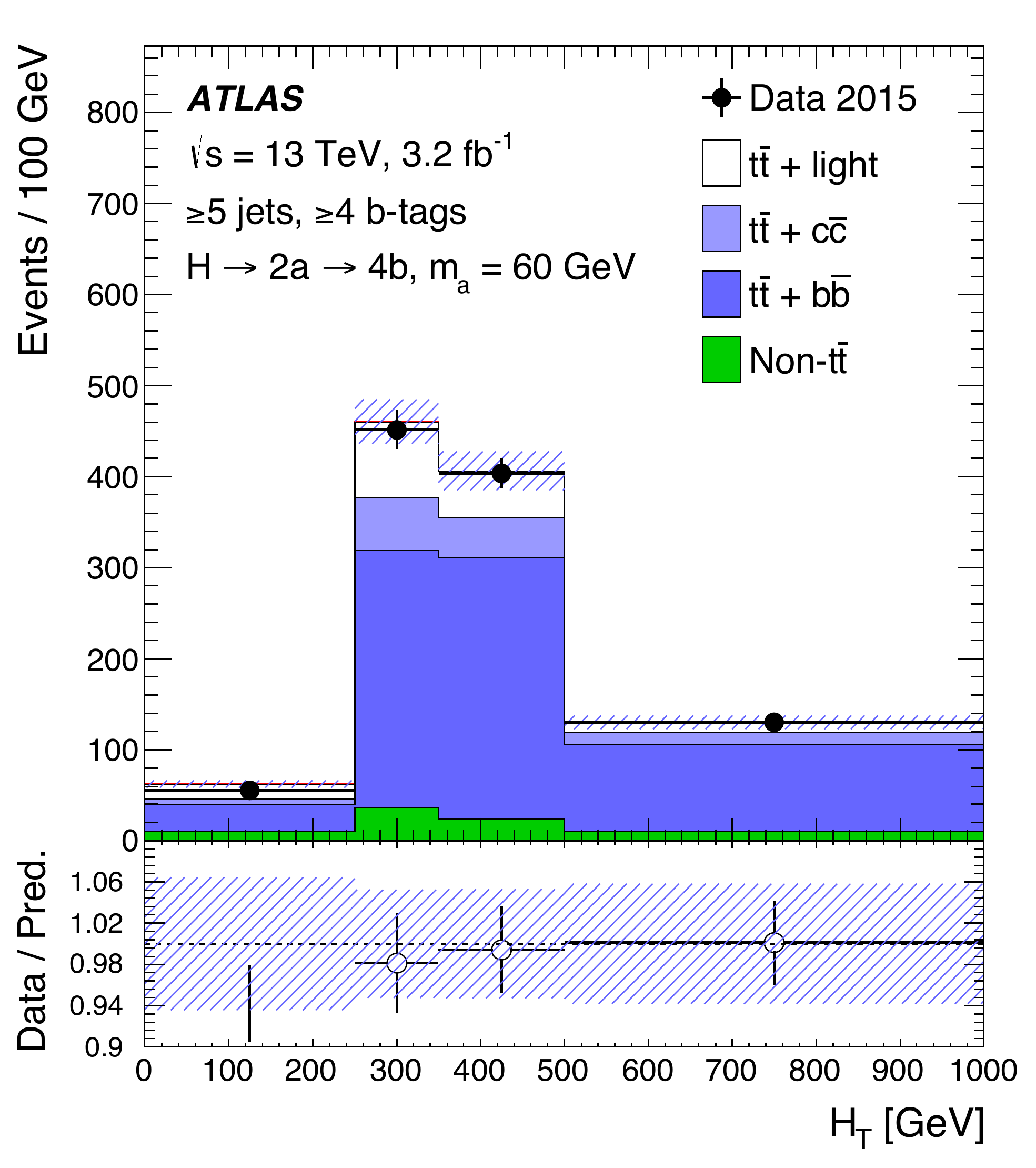}}
\caption{Comparison between the data and prediction for the
  distribution of the $H_{\rm T}$ variable used in the control regions
  with three and four $b$-tagged jets. 
  These distributions are after the fit is performed
  on data under the background-only hypothesis. The last bin contains the overflow.
}   
\label{fig:postfit_unblinded2}
\end{center}
\end{figure*}

\begin{figure*}[htbp]
\begin{center}
\subfloat[]{\includegraphics[width=0.45\textwidth]{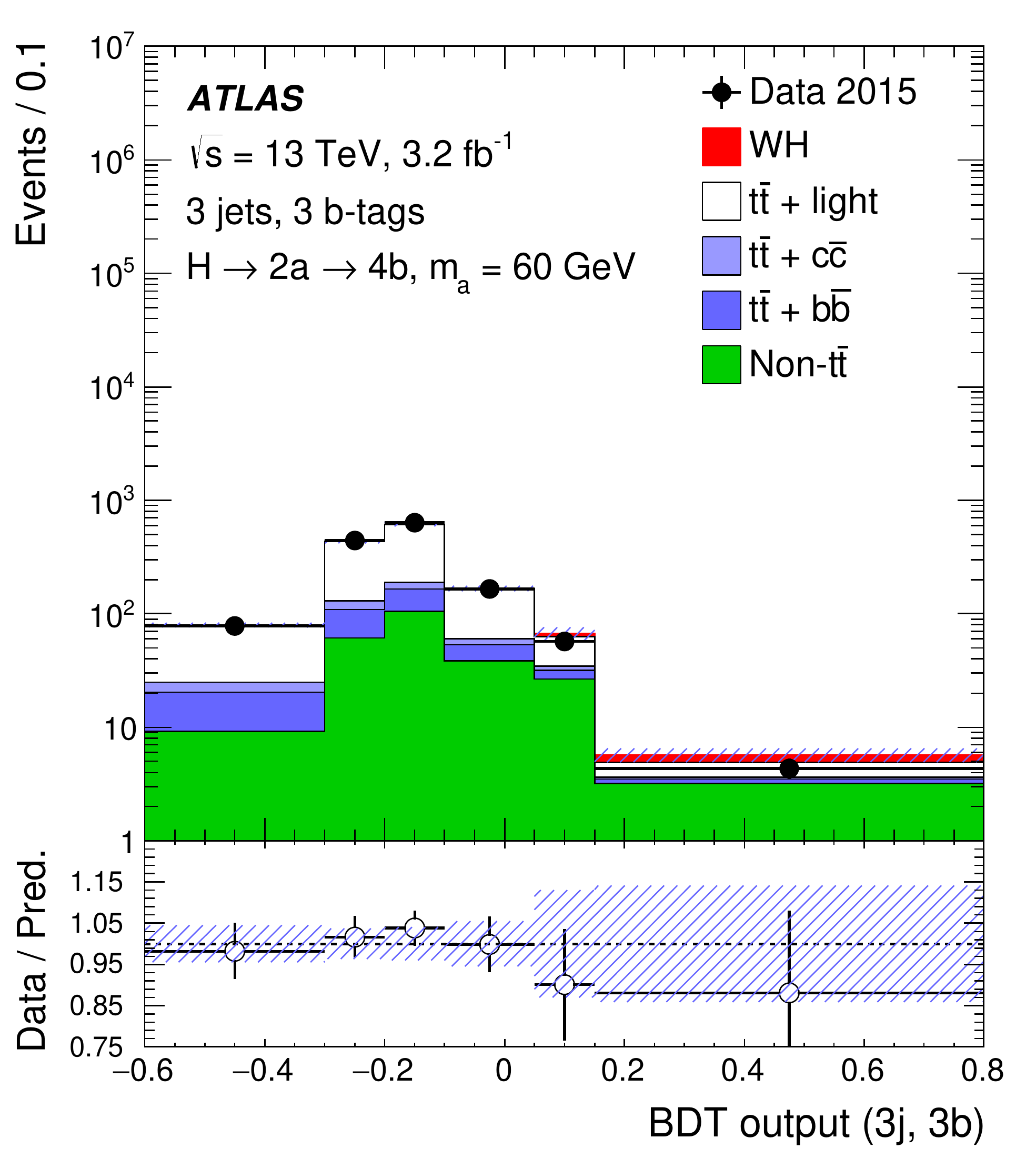}}%
\subfloat[]{\includegraphics[width=0.45\textwidth]{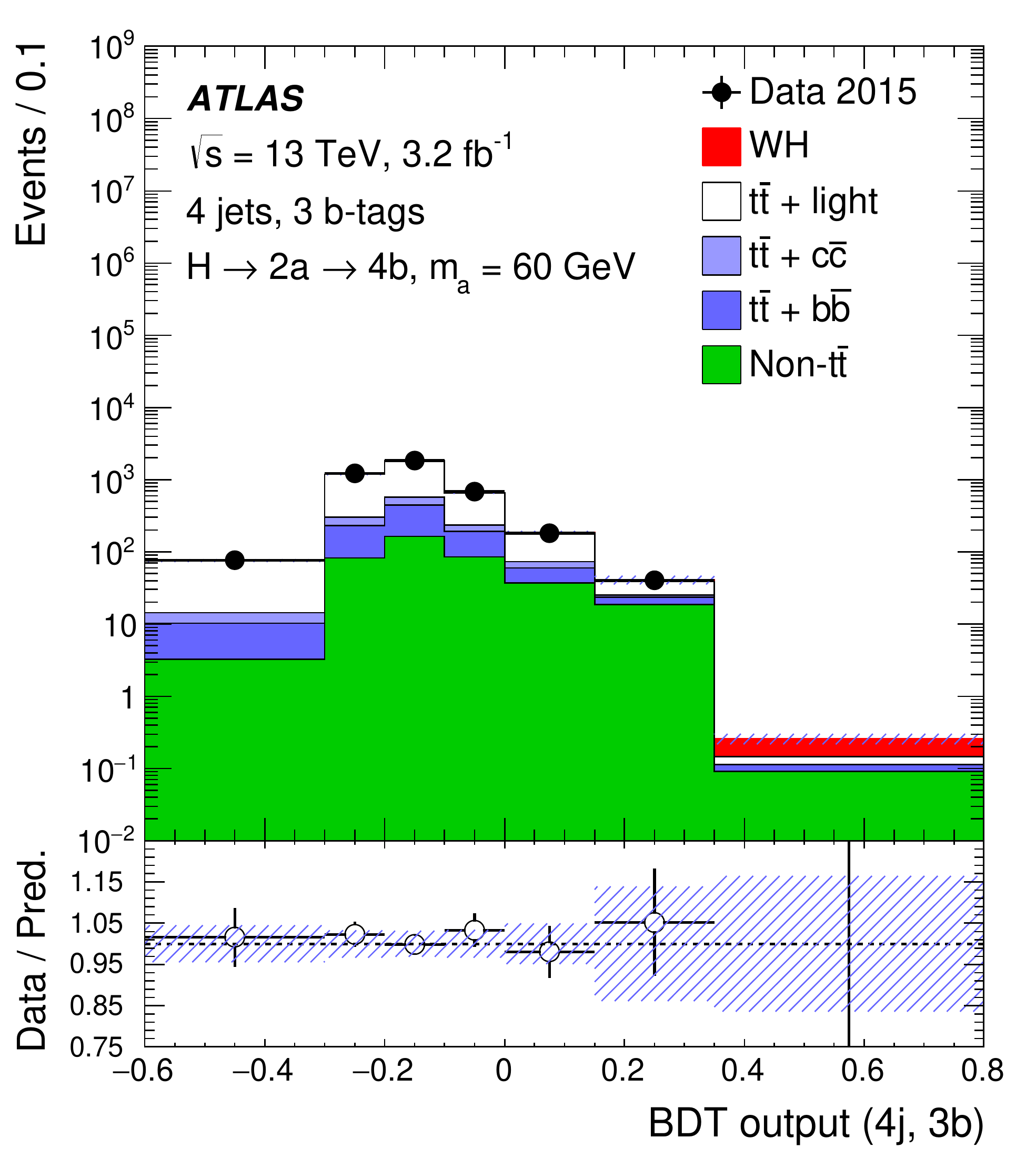}}\\%
\subfloat[]{\includegraphics[width=0.45\textwidth]{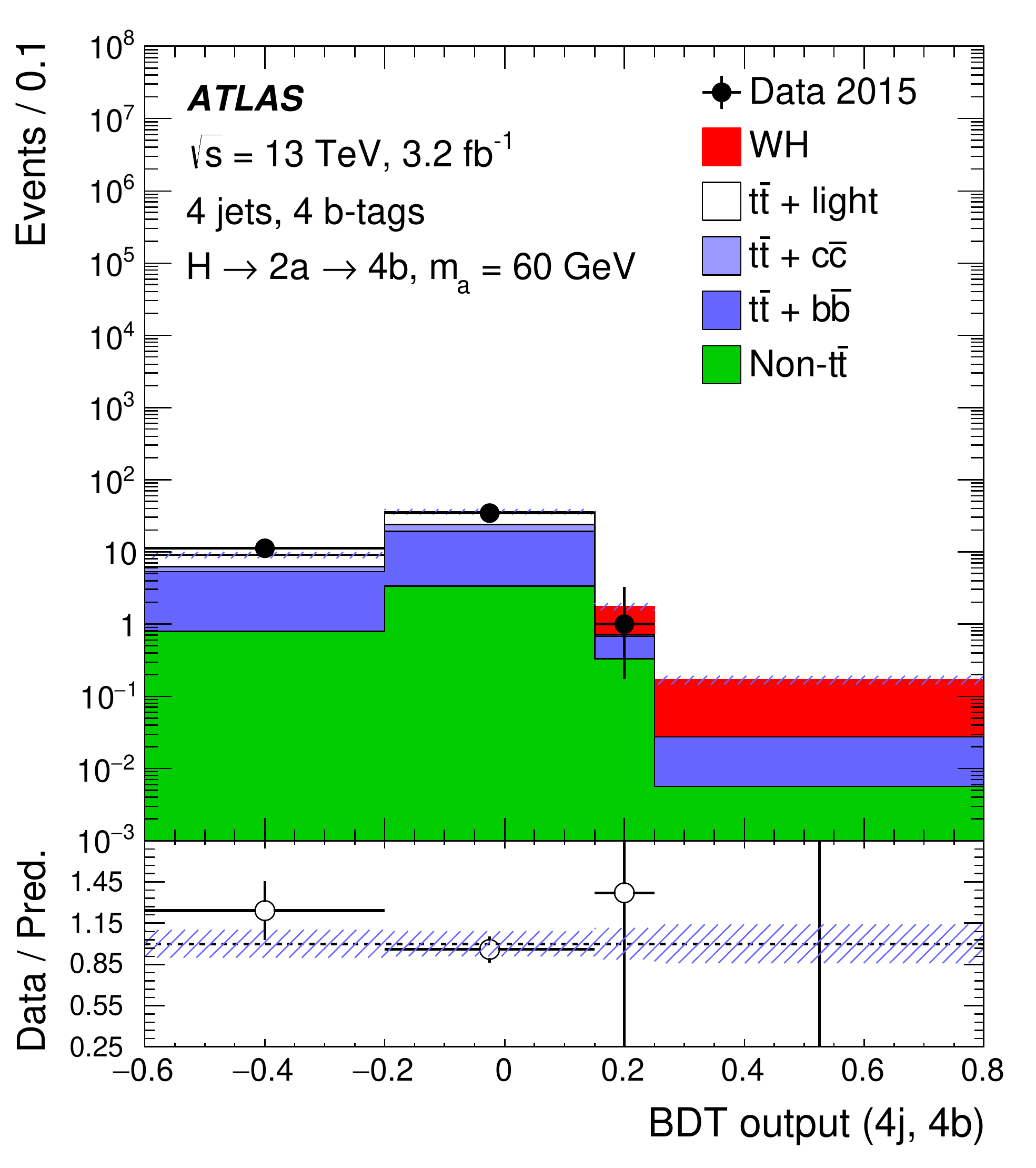}}
\caption{Comparison between the data and prediction for the
  distribution of the BDT discriminant used in the
  signal regions. These distributions are after the fit is performed
  on data under the background-only hypothesis. The hashed area 
  represents the total uncertainty in the background. 
  The distributions for the signal model ($WH$, $H\to 2a \to 4b$), 
  with $m_a=\unit[60]{\GeV}$, are normalised to the SM 
  $pp \rightarrow WH$ cross section, assuming 
  $\BR(H\to aa) \times \BR(a \to bb)^2 = 1$. The BDT output is
  determined in the range $[-1,1]$. The first and last bin
  contain the underflow and overflow, respectively. Markers are not
  drawn if they are outside the $y$-axis range.
}   
\label{fig:postfit_unblinded3}
\end{center}
\end{figure*}

The best fit of the background predictions to data in the binned
maximum-likelihood fit is shown in Figures~\ref{fig:postfit_unblinded1}--\ref{fig:postfit_unblinded3}.
Table~\ref{tab:signalregions_postfityields} shows the resulting 
yields and uncertainties for the signal regions after the fit. The SM
background yields obtained after performing the fit are in agreement
with the results from a fit using only the $\HT$ distributions in the
control regions. 

\begin{table}[h!]
\begin{center}
\begin{small}
  \begin{tabular*}{\textwidth}{@{\extracolsep{\fill}}lccc}
      \toprule
      Process & {(3j, 3b)} & {(4j, 3b)} & {(4j, 4b)} \\
      \hline
      $t\bar{t}$ + light                    & 1089 $\pm$ 76\hd         & 2940 $\pm$ 180           & 53 $\pm$ 16  \\
      $t\bar{t}$ + $c\bar{c}$               & \hd\hd70 $\pm$ 28\hd     & \hd280 $\pm$ 110         & 21 $\pm$ 11  \\
      $t\bar{t}$ + $b\bar{b}$               & \hd172 $\pm$ 55\hd       & \hd610 $\pm$ 160         & 74 $\pm$ 15  \\
      $t\bar{t}$ +$\gamma/\Wboson/\Zboson$  & \hd0.8 $\pm$ 0.1         & \hd\hd\hd4 $\pm$ 1\hd\hd & 0.4 $\pm$ 0.1  \\
      $W$ + jets                            & \hd\hd93 $\pm$ 31\hd     & \hd129 $\pm$ 40\hd       & 2 $\pm$ 1  \\
      $Z$ + jets                            & \hd\hd18 $\pm$ 12\hd     & \hd\hd14 $\pm$ 10\hd     & {--}  \\
      Single-top-quark                      & \hd135 $\pm$ 13\hd       & \hd208 $\pm$ 17\hd       & 8 $\pm$ 1  \\
      Multijet                              & \hd\hd48 $\pm$ 20\hd     & \hd\hd67 $\pm$ 28\hd     & 4 $\pm$ 2  \\
      Dibosons                              & \hd\hd\hd4 $\pm$ 1\hd\hd & \hd\hd\hd9 $\pm$ 1\hd\hd & 0.6 $\pm$ 0.4  \\
      $t\bar{t}$ + $H$                      & \hd0.7 $\pm$ 0.1         & \hd\hd\hd4 $\pm$ 1\hd\hd & 0.8 $\pm$ 0.2  \\
      \hline
      Total                                 & 1640 $\pm$ 58\hd         & 4270 $\pm$ 130 & 165 $\pm$ 15\hd  \\
      \hline
      Data                                  & 1646   & 4302 & 166  \\
      \hline
      $WH$, $H \to 2a \to 4b$   & \multicolumn{3}{r}{} \\
      $m_{a} = \unit[60]{\GeV}$   & \hd\hd10 $\pm$ 2\hd\hd  & \hd\hd\hd9 $\pm$ 1\hd\hd & \hd\hd\hd3 $\pm$ 1\hd\hd   \\
      $m_{a} = \unit[40]{\GeV}$   & \hd\hd11 $\pm$ 2\hd\hd  & \hd\hd10  $\pm$ 2\hd\hd & \hd\hd\hd2  $\pm$ 1\hd\hd  \\
      $m_{a} = \unit[20]{\GeV}$   & \hd\hd\hd6 $\pm$ 1\hd\hd      & \hd\hd\hd5 $\pm$ 1\hd\hd & \hd0.7 $\pm$ 0.2  \\
    \bottomrule
      \end{tabular*}
    \caption{\label{tab:signalregions_postfityields} Expected event
      yields of the SM background processes in the three signal
      regions after performing the fit with the background-only
      hypothesis. The observed data and the number of expected signal
      events are also indicated. The signal yields are quoted for some
      representative values of $m_{a}$ and assume the
      SM $pp \rightarrow WH$ cross section, $\sigma_{\rm SM}(WH) =
      1.37$~pb~\cite{Heinemeyer:2013tqa}, and $\BR(H\to aa) \times
      \BR(a \to bb)^2 = 1$. The uncertainties include statistical and
      systematic components (systematic uncertainties are discussed 
      in Section~\ref{sec:systematics}).
      The total uncertainty can differ from the sum in quadrature of
      individual sources due to correlations between them.
    }
  \end{small}
  \end{center}
  \end{table}

\begin{figure*}[htbp]
\begin{center}
\includegraphics[width=0.7\textwidth]{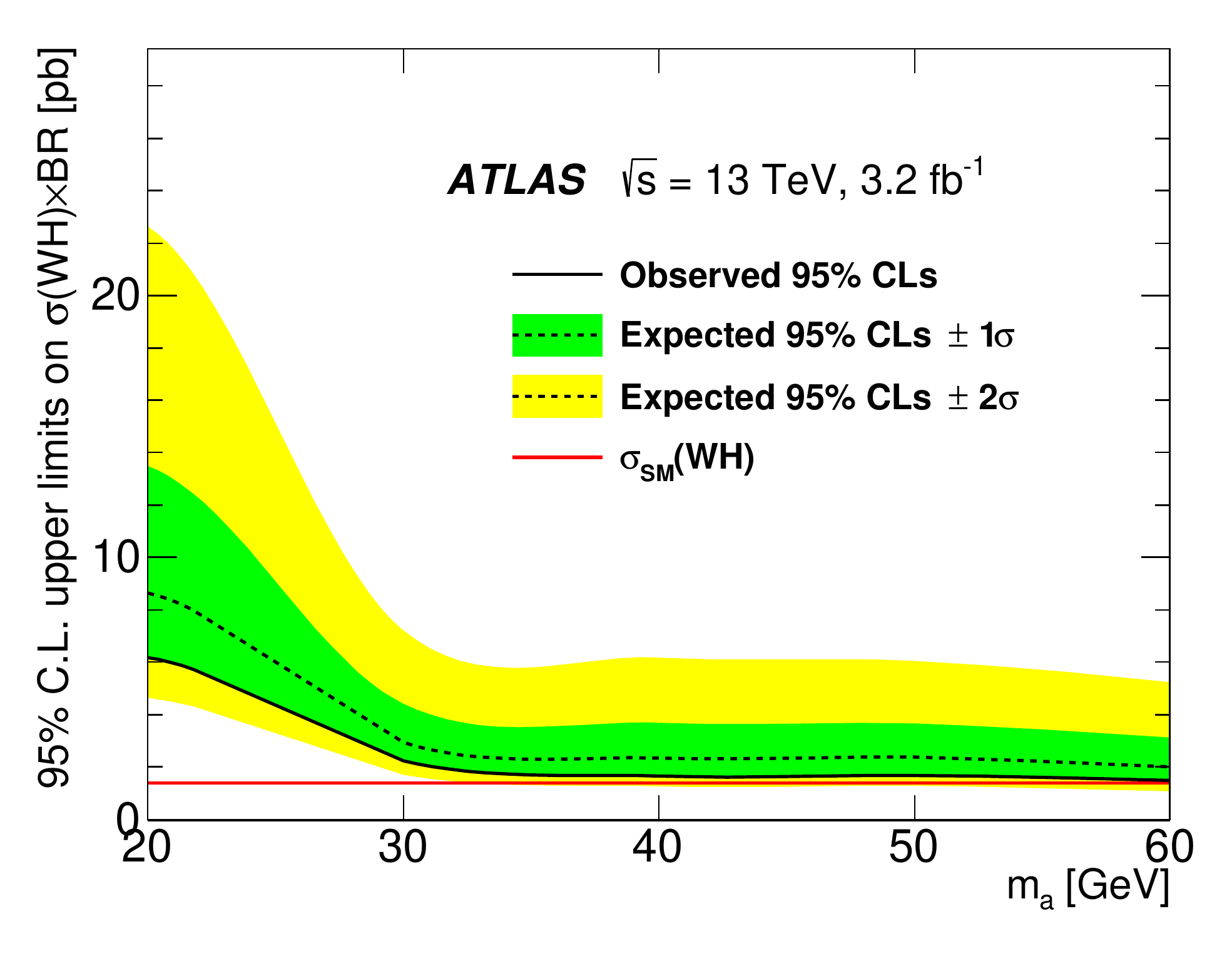}
\caption{Upper limit at 95\% CL on $\sigma(WH)\times \BR$, where $BR=\BR(H\to aa) \times \BR(a \to
  bb)^2$, versus $m_a$. The observed $({\mathrm{CL}}_{\mathrm{s}})$ values (solid black line)
  are compared to the expected (median) $({\mathrm{CL}}_{\mathrm{s}})$
  values under the background-only hypothesis (dotted black line). The surrounding 
  shaded bands correspond to the 68\% and 95\% CL intervals around the 
  expected $({\mathrm{CL}}_{\mathrm{s}})$ values, denoted by $\pm
  1\sigma$ and $\pm 2\sigma$, respectively. The solid red line
  indicates the SM $pp \to WH$ cross section, assuming 
  $\BR(H\to aa) \times \BR(a \to bb)^2 = 1$. Markers are not drawn if
  they are outside the $y$-axis range.}
\label{fig:CLs_vs_BR} 
\end{center}
\end{figure*}

In the absence of a significant excess of data above the background
prediction, upper limits are calculated for $\mu$, defined in
Eq.~(\ref{eq:br_mu}). The modified frequentist method ($({\mathrm{CL}}_{\mathrm{s}})$)~\cite{Read:2002hq} and
asymptotic formulae~\cite{Cowan:2010js} are used. Figure~\ref{fig:CLs_vs_BR}
shows the upper limits obtained at 95\% CL. The mass hypothesis
$m_a$ is tested in steps of $\unit[10]{\GeV}$ between $20$ and $\unit[60]{\GeV}$. 
The observed (expected) 95\% CL upper limits on $\mu$ range from $6.2~(8.6)$~pb, assuming 
$m_a = \unit[20]{\GeV}$, to $1.5~(2.0)$~pb, assuming 
$m_a = \unit[60]{\GeV}$. Assuming the SM $pp \rightarrow WH$ cross section, it
is not possible to set limits on the branching fraction with the
amount of data used. The reduced sensitivity for the light
$a$-boson hypothesis is due to a lower acceptance caused by overlapping
$b$-jets. The event yields indicated in Table~\ref{tab:signalregions_postfityields}
correspond to the sum of all BDT bins, while the fit
is most sensitive in the highest BDT bins, where the data are slightly
below the prediction, and hence the observed
limit is slightly lower than the expected one.

%
%
\section{Conclusion}
\label{sec:conclusion}

This paper presents a dedicated search for exotic decays of the Higgs boson to a
pair of new spin-zero particles, $H \rightarrow aa$, where the new
$a$-boson decays to $b$-quarks. The search focuses on the process $pp \to WH$ where 
the Higgs boson is produced in association with a $\Wboson$ boson. The 
analysis uses the $pp$ collision dataset at $\sqrt{s} = \unit[13]{\TeV}$
recorded by the ATLAS detector at the LHC in 2015, corresponding to an integrated
luminosity of $\unit[3.2 \pm 0.2]{\ifb}$. 
The search for $H \to 2a \to 4b$ is performed in the mass range 
$\unit[20]{\GeV} \leq m_a \leq \unit[60]{\GeV}$. The analysis uses several kinematic
variables combined in a multivariate discriminant in signal
regions and uses control regions to reduce the uncertainties in the 
backgrounds. No significant excess of data is observed relative to the
SM predictions.
Upper limits are derived for the product of the production cross section 
for $pp \to WH$ times the branching ratio for the decay $H \to 2a \to 4b$. 
The upper limit ranges from 6.2~pb for an $a$-boson mass $m_a =
\unit[20]{\GeV}$ to 1.5~pb for $m_a = \unit[60]{\GeV}$.

\section*{Acknowledgements}


We thank CERN for the very successful operation of the LHC, as well as the
support staff from our institutions without whom ATLAS could not be
operated efficiently.

We acknowledge the support of ANPCyT, Argentina; YerPhI, Armenia; ARC, Australia; BMWFW and FWF, Austria; ANAS, Azerbaijan; SSTC, Belarus; CNPq and FAPESP, Brazil; NSERC, NRC and CFI, Canada; CERN; CONICYT, Chile; CAS, MOST and NSFC, China; COLCIENCIAS, Colombia; MSMT CR, MPO CR and VSC CR, Czech Republic; DNRF and DNSRC, Denmark; IN2P3-CNRS, CEA-DSM/IRFU, France; GNSF, Georgia; BMBF, HGF, and MPG, Germany; GSRT, Greece; RGC, Hong Kong SAR, China; ISF, I-CORE and Benoziyo Center, Israel; INFN, Italy; MEXT and JSPS, Japan; CNRST, Morocco; FOM and NWO, Netherlands; RCN, Norway; MNiSW and NCN, Poland; FCT, Portugal; MNE/IFA, Romania; MES of Russia and NRC KI, Russian Federation; JINR; MESTD, Serbia; MSSR, Slovakia; ARRS and MIZ\v{S}, Slovenia; DST/NRF, South Africa; MINECO, Spain; SRC and Wallenberg Foundation, Sweden; SERI, SNSF and Cantons of Bern and Geneva, Switzerland; MOST, Taiwan; TAEK, Turkey; STFC, United Kingdom; DOE and NSF, United States of America. In addition, individual groups and members have received support from BCKDF, the Canada Council, CANARIE, CRC, Compute Canada, FQRNT, and the Ontario Innovation Trust, Canada; EPLANET, ERC, FP7, Horizon 2020 and Marie Sk{\l}odowska-Curie Actions, European Union; Investissements d'Avenir Labex and Idex, ANR, R{\'e}gion Auvergne and Fondation Partager le Savoir, France; DFG and AvH Foundation, Germany; Herakleitos, Thales and Aristeia programmes co-financed by EU-ESF and the Greek NSRF; BSF, GIF and Minerva, Israel; BRF, Norway; Generalitat de Catalunya, Generalitat Valenciana, Spain; the Royal Society and Leverhulme Trust, United Kingdom.

The crucial computing support from all WLCG partners is acknowledged gratefully, in particular from CERN, the ATLAS Tier-1 facilities at TRIUMF (Canada), NDGF (Denmark, Norway, Sweden), CC-IN2P3 (France), KIT/GridKA (Germany), INFN-CNAF (Italy), NL-T1 (Netherlands), PIC (Spain), ASGC (Taiwan), RAL (UK) and BNL (USA), the Tier-2 facilities worldwide and large non-WLCG resource providers. Major contributors of computing resources are listed in Ref.~\cite{ATL-GEN-PUB-2016-002}.

%

\printbibliography

\newpage 
\begin{flushleft}
{\Large The ATLAS Collaboration}

\bigskip

M.~Aaboud$^\textrm{\scriptsize 136d}$,
G.~Aad$^\textrm{\scriptsize 87}$,
B.~Abbott$^\textrm{\scriptsize 114}$,
J.~Abdallah$^\textrm{\scriptsize 65}$,
O.~Abdinov$^\textrm{\scriptsize 12}$,
B.~Abeloos$^\textrm{\scriptsize 118}$,
R.~Aben$^\textrm{\scriptsize 108}$,
O.S.~AbouZeid$^\textrm{\scriptsize 138}$,
N.L.~Abraham$^\textrm{\scriptsize 152}$,
H.~Abramowicz$^\textrm{\scriptsize 156}$
% ...
\bigskip
\\
% ...
$^{12}$ Institute of Physics, Azerbaijan Academy of Sciences, Baku, Azerbaijan\\
% ...
$^{65}$ University of Iowa, Iowa City IA, United States of America\\
% ...
$^{87}$ CPPM, Aix-Marseille Universit{\'e} and CNRS/IN2P3, Marseille, France\\
% ...
$^{108}$ Nikhef National Institute for Subatomic Physics and University of Amsterdam, Amsterdam, Netherlands\\
% ...
$^{114}$ Homer L. Dodge Department of Physics and Astronomy, University of Oklahoma, Norman OK, United States of America\\
% ...
$^{118}$ LAL, Univ. Paris-Sud, CNRS/IN2P3, Universit{\'e} Paris-Saclay, Orsay, France\\
% ...
$^{136}$ $^{(a)}$ Facult{\'e} des Sciences Ain Chock, R{\'e}seau Universitaire de Physique des Hautes Energies - Universit{\'e} Hassan II, Casablanca; $^{(b)}$ Centre National de l'Energie des Sciences Techniques Nucleaires, Rabat; $^{(c)}$ Facult{\'e} des Sciences Semlalia, Universit{\'e} Cadi Ayyad, LPHEA-Marrakech; $^{(d)}$ Facult{\'e} des Sciences, Universit{\'e} Mohamed Premier and LPTPM, Oujda; $^{(e)}$ Facult{\'e} des sciences, Universit{\'e} Mohammed V, Rabat, Morocco\\
% ...
$^{138}$ Santa Cruz Institute for Particle Physics, University of California Santa Cruz, Santa Cruz CA, United States of America\\
% ...
$^{152}$ Department of Physics and Astronomy, University of Sussex, Brighton, United Kingdom\\
% ...
$^{156}$ Raymond and Beverly Sackler School of Physics and Astronomy, Tel Aviv University, Tel Aviv, Israel\\
% ...
$^{a}$ Also at Department of Physics, King's College London, London, United Kingdom\\
$^{b}$ Also at Institute of Physics, Azerbaijan Academy of Sciences, Baku, Azerbaijan\\
$^{c}$ Also at Novosibirsk State University, Novosibirsk, Russia\\
$^{d}$ Also at TRIUMF, Vancouver BC, Canada\\
$^{e}$ Also at Department of Physics {\&} Astronomy, University of Louisville, Louisville, KY, United States of America\\
% ...
\end{flushleft}




%

\end{document}